\documentclass[prb,twocolumn]{revtex4}
\usepackage[dvips]{graphicx}
\usepackage{latexsym}
\usepackage{amsmath, bm, eufrak, mathrsfs}
\usepackage{wasysym}

\def\etal{~\textit{et~al.}} 
\def\ra{\rangle} 
\def\la{\langle} 
\def\const{{\rm const}}

\def\g{{\EuFrak g}}
\def\h{{\EuFrak h}}
\def\bmnabla{{\bm \nabla}}
\def\Cscr{{\mathscr C}}
\def\bma{{\bm a}}
\def\bmc{{\bm c}}
\def\bme{{\bm e}}
\def\bmA{{\bm A}}
\def\bml{{\bm l}}

\def\mod{\rm{mod}}

\begin{document}

\title{On the origin of artificial electrodynamics and other stories 
in three-dimensional bosonic models}
\author{O. I. Motrunich}
\affiliation{Kavli Institute for Theoretical Physics,
University of California, Santa Barbara, CA 93106-4030}
\author{T. Senthil}
\affiliation{Massachusetts Institute of Technology, 
77 Massachusetts Ave., Cambridge, MA 02139-4307}

\date{\today}

\begin{abstract}
Several simple models of strongly correlated bosons on 
three-dimensional lattices have been shown to possess exotic 
fractionalized Mott insulating phases with a gapless `photon' 
excitation.  In this paper we show how to view the physics of this 
`Coulomb' state in terms of the excitations of proximate superfluid.
We argue for the presence of ordered vortex cores with a broken 
discrete symmetry in the nearby superfluid phase, and that 
proliferating these degenerate but distinct vortices with equal 
amplitudes produces the Coulomb phase.
This provides a simple physical description of the origin of the 
exotic excitations of the Coulomb state.
The physical picture is formalized by means of a dual description
of three-dimensional bosonic systems in terms of fluctuating 
quantum mechanical vortex loops.
Such a dual formulation is extensively developed.  It is shown how the 
Coulomb phase (as well as various other familiar phases) of 
three-dimensional bosonic systems may be described in this vortex loop 
theory.  For bosons at half-filling and the closely related system of 
spin-$1/2$ quantum magnets on a cubic lattice, fractionalized phases as 
well as bond or `box' ordered states are possible. 
The latter are analyzed by an extension of techniques previously 
developed in two spatial dimensions.  The relation between these 
`confining' phases with broken translational symmetry and the 
fractionalized Coulomb phase is exposed.
\end{abstract}

\maketitle

\section{Introduction}
Several recent studies have produced a variety of models that exhibit 
quantum number fractionalization in two dimensions and in the absence 
of magnetic fields.\cite{RSSpN, MoeSon2D, Iof, BalMPAFGir, bosfrc}
It has been appreciated for some time that similar fractionalization 
phenomenon can occur also in three dimensions, and can take novel forms 
not possible in 2D.\cite{bosfrc, Wen, Huse, Hermele, MoeSon3D}
In particular, so called Coulomb phases have recently been demonstrated 
in 3D bosonic models.  In these Coulomb phases the fractionalized 
degrees of freedom interact with a gapless emergent $U(1)$ gauge field.  
The gauge excitations describe, at low energies and long distances, 
a linear dispersing transverse mode which has been dubbed an 
`artificial photon'.  The adjective `artificial' refers to the fact that 
this photon really corresponds to some collective excitation of the 
boson system and has nothing directly to do with the true 
electromagnetic photon. 
In particular the `artificial photon' in the fractionalized 3D Coulomb 
phases arises even in microscopic models of correlated bosons 
with purely short-ranged interactions.  The Coulomb phase also contains 
additional gapped excitations that may be identified with 
magnetic monopoles of the internal gauge field.
The possibility of such dynamical generation of `light' was noted more
than two decades ago in Ref.~\onlinecite{Nielsen}.

In the case of fractionalization in 2D bosonic systems, a very 
fruitful perspective is obtained by departing from a superfluid 
state and considering nearby Mott insulator phases as vortex 
condensates.\cite{MPAFdual, NLII, z2long, Lannert}
Conventional insulating states of bosons, with possibly
charge or bond order, can be understood in these terms.
The same perspective also naturally predicts occurrence of
a $Z_2$ fractionalized state when the superfluidity is destroyed 
by the condensation of doubled vortices while single ones remain gapped. 
The unpaired single vortex survives in the resulting insulating phase 
as a gapped excitation that retains a $Z_2$ quantum number 
(only the oddity of vorticity remains well defined) --- this is the 
$Z_2$ vortex or vison excitation of this fractionalized insulator. 

Turning to 3D systems, we envisage a similar perspective on 
insulating states as vortex condensates, except that vortex
excitations in the superfluid are now lines instead of point particles.
For example, a $Z_2$ fractionalized insulator is obtained
by destroying the superfluid order by pairing and condensing 
vortex lines.

The goal of the present paper is to develop a similar intuition
for the 3D Coulomb phase (also dubbed $U(1)$ phase).  
In particular, we want to understand the genesis of the excitation 
structure of this phase in an approach that departs from the superfluid 
state.  We note that the excitation structure of the Coulomb phase 
consists of gapped charge-$1/2$ chargons that are minimally coupled to 
an emergent gapless ``photon'' mode.  In addition, there is another 
gapped excitation that may be viewed as a ``magnetic monopole'' that 
acts as a source of the emergent magnetic flux.  The picture that 
appears from the present study is that physics inside vortex cores is 
intimately involved in producing the $U(1)$ phase.  We thus have a 
rather unusual but very interesting situation when we cannot ignore the 
microscopics of the vortex core when describing the relationship of the
superfluid phase to the nearby insulating states.

\begin{figure}
\centerline{\includegraphics[width=\columnwidth]{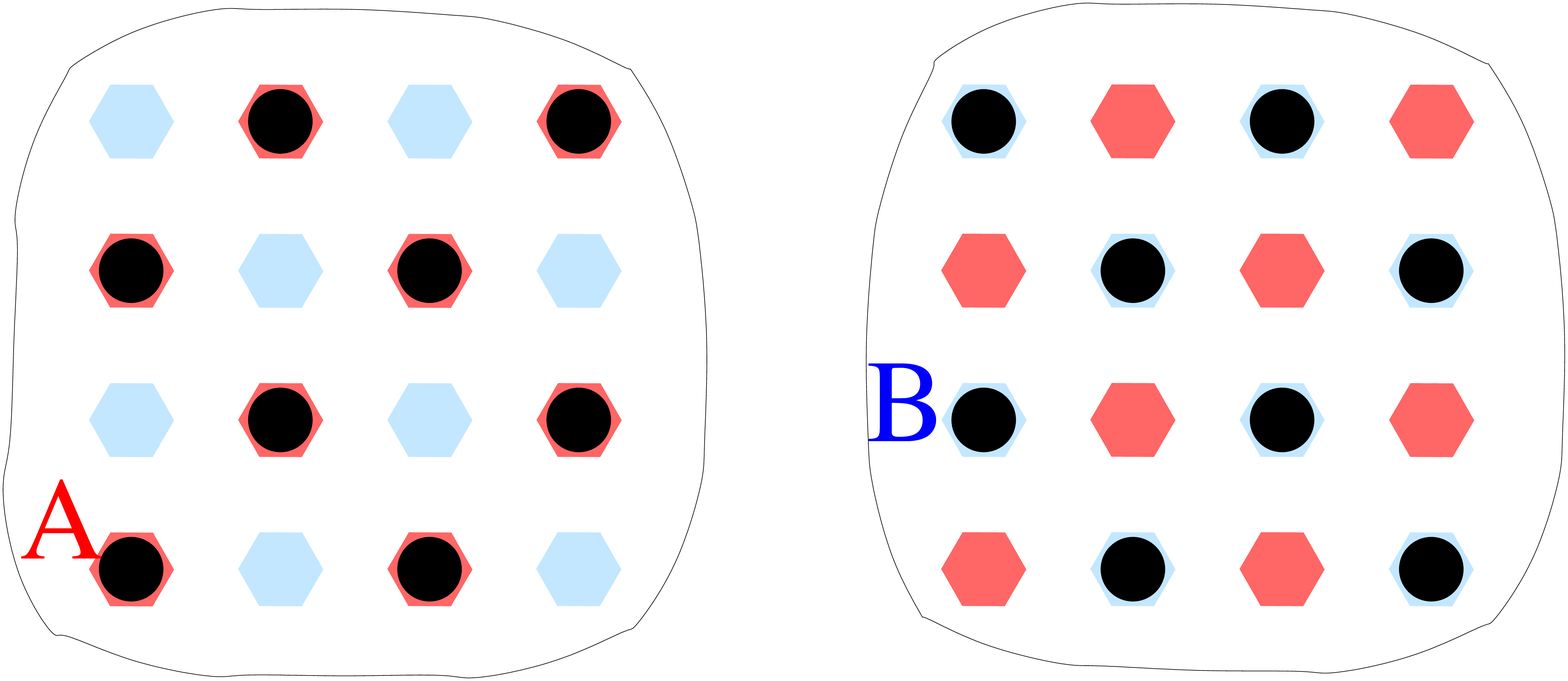}}
\caption{View of the cross section of a vortex line in a 
three-dimensional superfluid of bosons at half-filling on a 
cubic lattice.  Two checkerboard ordering patterns --- where 
the boson density is higher on one of the two sublattices --- are 
possible inside the vortex core.
}
\label{fig:cdw}
\vskip 2mm
\centerline{\includegraphics[width=\columnwidth]{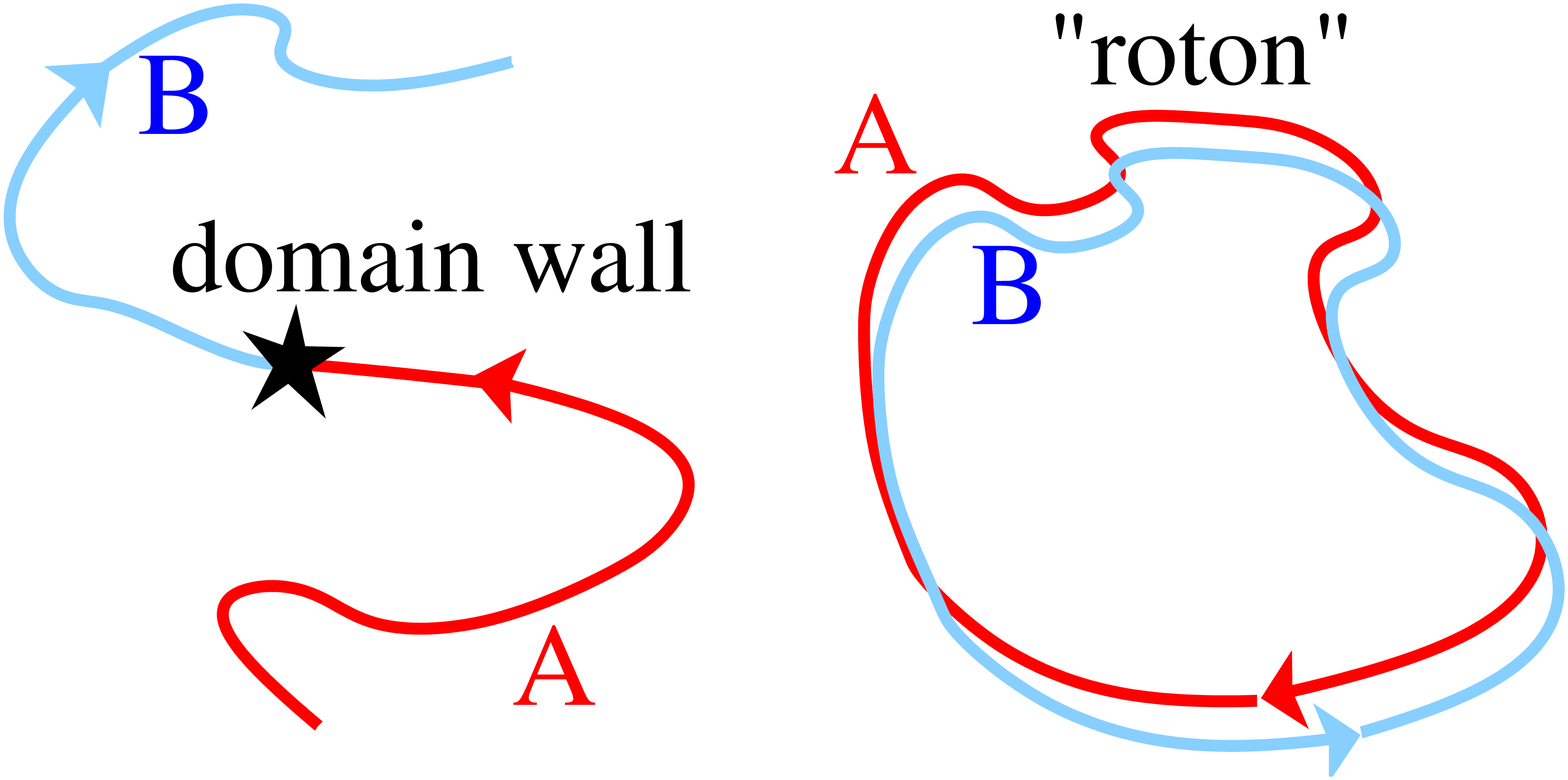}}
\caption{Vortex lines with different core orders are indicated
as $A$ (red) and $B$ (light blue). 
Left panel:  A domain wall associated with the order in the core.  
If the superfluid is disordered by proliferating both vortex species
simultaneously while keeping such domain walls gapped, the Coulomb 
phase results.  The gapped domain walls survive as the monopoles of the 
Coulomb phase.
Right panel:  A ``roton'' loop is formed by bringing together a vortex 
with one ordered core and an antivortex with the other core. 
These particular roton loops correspond to loops of emergent magnetic 
flux which are strongly fluctuating in the Coulomb phase.
The artificial photon is associated with fluctuations of these loops. 
}
\label{fig:mnpl_roton}
\end{figure}

The main idea is very simply illustrated by considering bosons on a 
three-dimensional cubic lattice and at an average filling of one half 
boson per site.  In some semiclassical description, 
there are two simple phases that might be imagined for the bosons. 
First, there is a translationally invariant superfluid phase. 
Second, there is a Mott insulator in which the bosons preferentially 
occupy sites on one or the other of the two sublattices. 
Clearly, such a Mott insulating ground state is two-fold degenerate
corresponding to the two possible states of boson charge ordering.  
The superfluid phase possesses vortices which are extended line defects. 
The superfluid order parameter will be suppressed in the core of
such a vortex line.  If the vortex core size is larger than the 
lattice spacing (as might well happen) it makes sense to ask about what 
`phase' is obtained in the core.  An immediate guess is that the core is 
simply the checkerboard charge-ordered Mott insulating state as
illustrated in Fig.~\ref{fig:cdw}. 
However the checkerboard state breaks a symmetry (that of sublattice 
interchange), and is two-fold degenerate. 
We are then led to conclude that there will be two kinds of vortices 
depending on which sublattice has higher boson density in the core. 

This simple-minded discussion on the structure of vortices will be 
modified once we take quantum fluctuations into account more seriously 
in the vortex core.  In particular, as the core has finite radius,
it essentially behaves as a one dimensional system. 
Quantum fluctuations will tend to restore the broken symmetry in the 
core and produce a unique vortex line.  However, as the broken sublattice
symmetry is discrete, it is natural to expect that the symmetry breaking 
will be stable to weak quantum fluctuations.  Thus there will be some 
range of parameters in which there will be two distinct kinds of vortex 
lines with symmetry broken cores. 
In other range of parameters a unique vortex with no symmetry breaking 
in the core may well result.  The change between the two parameter ranges
involves no ground state phase transition but only a change in the 
nature of the excited states. 

Now consider disordering the superfluid by proliferating and 
condensing vortex line loops.  Clearly, the insulating phase that 
obtains will depend on the structure of the vortices that are actually 
proliferating.  Consider the situation where the vortices have ordered 
cores.  If we preferentially condense one or the other kind of vortex, 
we will clearly induce checkerboard charge order in the insulator. 
But what if we condense both species of vortices with equal amplitude? 
We argue that the resulting phase is the Coulomb fractionalized 
insulator.  We show how the excitations of this exotic insulator may be 
understood in terms of the various excitations of the superfluid 
when it has ordered vortex cores. 

Consider a vortex line.  It is still possible for the core order
to change from one state to the other somewhere along the line
as shown in Fig.~\ref{fig:mnpl_roton}, 
but there is a gap for such a domain wall excitation.
We expect this gap to remain when such vortices proliferate while 
retaining their order.  When both vortex species proliferate,
the domain wall excitations become point particles which we identify
with the monopoles of the Coulomb phase.

The appearance of the artificial photon is more subtle, but is 
also a consequence of proliferating two species of vortices. 
To explain this, it is useful to recall what happens in the more
familiar situation of particle condensates.  When the particle that is 
condensing is `neutral' (in the sense of having only short ranged 
interactions), the resulting condensate supports gapless Goldstone 
excitations.  On the other hand, if it is charged (in the sense of 
having long ranged interactions), there are no gapless excitations 
associated with the condensation (the Anderson-Higgs phenomenon). 
Consider now the dual description of bosonic Mott insulators in 
three spatial dimensions as `condensates' of extended vortex loops. 
For such a line condensate we may again expect that if the vortices 
have long ranged interactions then the resulting condensate supports 
only gapped excitations.  However, consider now the situation envisaged 
above where there are two vortex species (distinguished by the ordering 
in the core) which both condense.  We may form a `roton' line from these 
two species by bringing together a vortex with one ordered core and an 
antivortex with the other core, and this is pictured in 
Fig.~\ref{fig:mnpl_roton}.  Such objects have no net vorticity, 
and therefore only short ranged interactions with one another. 
When both vortex species condense such roton loops condense as well. 
By analogy to particle condensates we may expect then to find gapless 
excitations.  We will show in this paper that this is indeed the case, 
and that the resulting gapless modes may be identified with the photon 
of the Coulomb phase.  
More precisely, we will argue that the roton lines described above may 
be viewed as the magnetic field lines of the emergent deconfined $U(1)$ 
gauge field that obtains in the Coulomb phase.

It is useful to point a connection with a recent study\cite{deccp} 
of so called deconfined critical points in two dimensions such as 
separating a superfluid and a bond density wave insulator of bosons at 
half-filling on a square lattice.  In 2D, a vortex is a point particle, 
and because of the quantum tunneling events we expect a unique physical 
vortex in the superfluid phase.  In other words, generically the two 
different ordering patterns in the core will tunnel into each other at a 
finite rate in $d = 2$.  However, at the critical point at which 
vortices proliferate, the tunneling becomes irrelevant.\cite{deccp} 
We may then effectively speak of two vortex species.  
The tunneling becomes again relevant in the vortex condensate, and 
produces a particular insulating state which in this example has bond 
density wave order.  However, as elaborated in Ref.~\onlinecite{deccp} 
there is a large intermediate region of length scales in the insulator 
(close to the critical point) which may be described as a two-dimensional
version of the Coulomb phase.  The (intermediate) long wavelength physics
of this region is again described in terms of a gapless excitation which 
may be regarded as the two-dimensional version of the photon.  
In three dimensions, having an ordered vortex core is a stable 
possibility in the superfluid.  As mentioned in the previous paragraph, 
proliferating both vortex species with equal amplitude then leads to a 
stable insulating phase which has a gapless photon excitation --- 
precisely the Coulomb state at the focus of the present study.

The primary purpose of this paper is to develop the physical picture 
sketched above for the excitations of the Coulomb phase in some detail. 
This goal is strongly aided by a formal dual description of 
three-dimensional bosonic systems in terms of extended fluctuating 
vortex loops.  This duality transformation --- a generalization to 
(3+1)D of the familiar boson-vortex duality in two spatial 
dimensions --- has been available in the literature for many years, 
and is reviewed in Section~\ref{sec:dual}.  
Here we will use this and further develop it extensively to obtain a 
dual description of the phases of various three-dimensional bosonic 
systems which puts meat into the physical pictures.  
In the course of these studies, we develop a description of bosons 
at half-filling on a three-dimensional cubic lattice that is capable of 
describing all the insulating phases discussed above --- the Coulomb 
phase, the checkerboard charge ordered phase, and also different bond 
density wave ordered phases.
This gives a three-dimensional generalization of the corresponding
development in Ref.~\onlinecite{deccp}.  Boson systems at half-filling 
are closely related to spin-$1/2$ quantum magnets.  Thus some of our 
results have direct implications for quantum spin systems on cubic 
lattices as well.  Indeed we will borrow from techniques familiar 
from studies of quantum antiferromagnets in two spatial dimensions, 
and generalize them in discussing the structure of the bond ordered 
insulating phases of bosons at half-filling

The paper is organized as follows.  In Section~\ref{sec:real}, 
we first review a simple bosonic model that explicitly realizes the 
fractionalized Coulomb phase.  This provides a good starting point for 
the discussion of the relationship of the $U(1)$ state to other phases.  
The model has integer boson filling and is simple to analyze in the 
charge representation, but the vortex physics is obscured by the 
nontrivial lattice structure.  
Still, the model motivates an effective theory, formulated as a 
compact $U(1)$ gauge theory with two matter fields (chargons), 
that properly captures the possible phases and is amenable to an 
analytic description in terms of vortices, furnishing the main argument 
for the proposed physical picture.  In Section~\ref{sec:vortex_phenom}, 
we consider vortex phenomenology in this two chargon theory.  
Precise duality analysis is presented in Section~\ref{sec:dual}, 
where we first review the conventional boson-vortex duality in (3+1)D 
from a modern perspective, and then describe how to include the physics 
of two vortex species in the dual description (the details of the 
derivation are given in Appendix~\ref{app:2ch}).
In Appendix~\ref{app:Ising}, we consider a (2+1)D Ising version of the 
specific boson model in order to give some intuition in a more simple 
setting; this toy model realizes a topological $Z_2$ phase in two
dimensions via a mechanism that resembles the one proposed for the 
Coulomb phase in three dimensions, with the vortices replaced by
Ising domain walls.
In Section~\ref{sec:halfint} and supporting 
Appendices~\ref{app:halfint}~and~\ref{app:2ch_halfint},
we develop a dual description for bosons at half-integer filling.
An important ingredient of this description is monopole Berry phases
generalizing Haldane's result to three-dimensional spin-1/2 quantum
magnets on a cubic lattice.  We apply this in Section~\ref{sec:vbs},
where we show how bond density wave states can be analyzed as monopole 
condensates coming from the Coulomb phase.

\section{Realization of the Coulomb phase and phenomenology}
\label{sec:real}
\subsection{Review of boson Hubbard model}
\begin{figure}
\centerline{\includegraphics[width=3.0in]{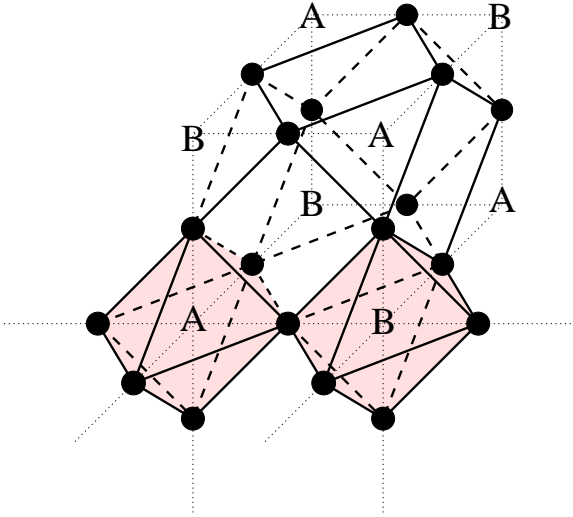}}
\vskip -2mm
\caption{Explicit model that realizes Coulomb phase.
Boson islands are located on the link-centered sites
of a simple cubic lattice.  The islands can be also viewed as
forming a network of corner-sharing octahedra, and two octahedron
units are shown.
In the Hamiltonian Eq.~(\ref{Hboson}), we stipulate a term 
$U N_r^2$ which prefers charge neutrality of each octahedron.
}
\label{lattice}
\end{figure}

We begin with a review of a specific model that realizes 
Coulomb phase in 3D.\cite{bosfrc, Wen, Hermele}
The model is formulated in terms of soft-core bosons
(quantum rotors) residing on link-centered sites (``islands'') 
of a simple cubic lattice as depicted in Fig.~\ref{lattice}.
The boson islands can also be viewed as forming a lattice of
corner-sharing octahedra.
Bosons can hop with amplitude $w$ between neighboring islands.
Besides the usual on-site repulsion $u_\psi$, we also include
repulsive interactions between the bosons that favor charge 
neutrality of each octahedron.
The complete Hamiltonian is
\begin{eqnarray*}
\label{Hboson}
H & = &
-w \sum_{\la ll' \ra}(\psi_{l}^\dagger \psi_{l'} + h.c.) 
+ u_\psi \sum_{l} n_l^2 
+ U \sum_r N_r^2 ~,
\end{eqnarray*}
where $\psi_l^\dagger = e^{i\phi_l}$ creates a boson on a given
island and $n_l$ is the corresponding boson number operator.
Each octahedron is labeled by the cubic lattice site $r$
at its center; the corresponding operator $N_r$ is defined 
through 
\begin{equation}
N_r = \sum_{l \in r} n_l ~,
\end{equation}
where $l \in r$ sums over all cubic lattice links emanating from $r$.  
The total boson number of the system is 
$N_{\rm tot} = \frac{1}{2} \sum_r N_r$.

\begin{figure}
\centerline{\includegraphics[width=3.0in]{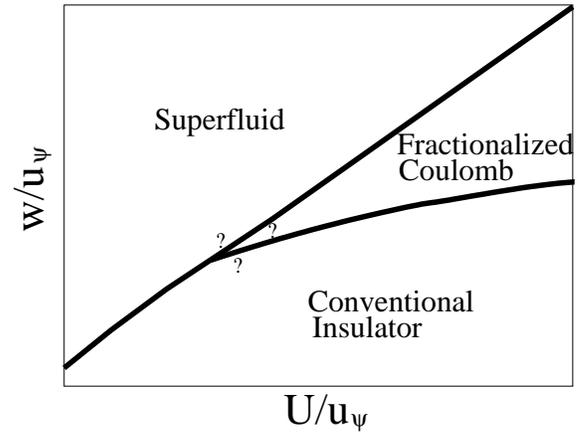}}
\vskip -2mm
\caption{Schematic phase diagram of the boson model Eq.~(\ref{Hboson}) 
exhibiting the stable phases discussed in the text;
note that we do not know the details for intermediate 
coupling strengths and whether some other states may intervene.
}
\label{phased}
\end{figure}

We summarize the analysis of the possible phases of the 
above Hamiltonian. 
When the boson hopping dominates, $w \gg u_\psi, U$, 
the system is a superfluid.  When the charging energy dominates, 
$u_\psi, U \gg w$, the system is a conventional Mott insulator.
On the other hand, as reviewed below, when the charging energies 
$U$ and $u_\psi$ are varied separately, there is an intermediate regime 
$U \gg w \gg \sqrt{u_\psi U}$ such that the system is in a
Coulomb phase\cite{bosfrc}.  This phase is a fractionalized insulator 
with the excitation spectrum consisting of gapped charge-1/2 chargons,
a gapless linearly dispersing photon, and a gapped monopole.
A schematic phase diagram of our model is shown in Fig.~\ref{phased}.

The analysis in the limit $U \gg w, u_\psi$ is similar to that in the 
large $U$ limit of the electronic Hubbard model at half-filling.
For $w=u_\psi=0$ there is a degenerate manifold of ground states 
specified by the requirement $N_r=0$ for each $r$, separated by a large 
charge gap $U$ from the nearest sectors.
Including the $w, u_\psi$ terms lifts the degeneracy in each such
sector, and this is best described by deriving the corresponding 
effective Hamiltonians for small perturbing couplings.  

In the ground state sector $N_r = 0$ for all $r$,
an elementary calculation gives 
\begin{eqnarray}
\label{Heff}
H_{\rm eff}^{(0)} = u_\psi \sum_{l} n_l^2 
-K_{\rm ring} \sum_\Box 
  \left( \psi_{12}^\dagger \psi_{23} \psi_{34}^\dagger \psi_{41} 
         + h.c. \right)
\end{eqnarray}
with $K_{\rm ring} = 2 w^2/U$.
When writing the `ring exchange' term around a given placket,
we label a boson island by the corresponding link end-points.

A simple change of variables shows that $H_{\rm eff}^{(0)}$ 
together with the constraint $N_r = 0$ can be regarded as the 
well-studied (3+1)D compact $U(1)$ gauge theory.
Indeed, divide the underlying cubic lattice of Fig.~\ref{lattice}
into $A$ and $B$ sublattices.  We now define a vector field 
$a_{r\alpha} \equiv \eta_r \phi_{r,r+\alpha}$, where
$\eta_r = +1$ if $r\in A$ and $\eta_r = -1$ if $r\in B$,
$\alpha = \hat{x}, \hat{y}, \hat{z}$;
we also perform the corresponding transformation to the vector
field $E_{r\alpha} \equiv \eta_r n_{r,r+\alpha}$ conjugate
to $a_{r\alpha}$.  In the new variables
\begin{equation}
\label{Nr}
N_r = \eta_r \bmnabla \cdot {\bm E} ~.
\end{equation}
If $a_{r\alpha}$ is interpreted as a compact $U(1)$ vector 
potential and $E_{r\alpha}$ as the corresponding electric
field operator, the ground state sector constraint $N_r=0$ 
is simply the Gauss law, while the effective Hamiltonian has the
declared lattice gauge theory form
\begin{equation}
\label{Heff0}
H_{\rm eff}^{(0)} =
u_\psi \sum_{r, \alpha} E_{r\alpha}^2
- 2 K_{\rm ring}\sum_\Box \cos(\bmnabla \times \bma) ~.
\end{equation}

In (3+1)D, the compact $U(1)$ gauge theory has two distinct
phases:  For $K_{\rm ring} \lesssim u_\psi$, the
gauge theory is confining, and all excitations carrying
non-zero gauge charge are confined.
Zero gauge charge excitations of course exist with a gap $\propto U$;
these carry integer physical charge in units of elementary
boson charge $q_b$.  
This is the conventional Mott insulator of our boson model.

In the opposite regime, $K_{\rm ring} \gtrsim u_\psi$,
the gauge theory is in the deconfined Coulomb phase,
and has a gapless linearly dispersing gauge boson (photon) and 
a gapped topological point defect (monopole) as its distinct 
excitations in the $N_r=0$ sector.
In the charged sectors, objects with $N_r = 1$ at some site, 
{\em i.e.}~physical charge $\frac{q_b}{2}$,
are not confined and can propagate above a finite gap of order $U$.
These charged excitations interact via an emergent long-range Coulomb 
interaction.  A detailed description of the ground state properties
of the Coulomb phase due to the gapless photon can be found in 
Ref.~\onlinecite{Hermele}.

This completes the review part of our discussion, and we now
focus on the relationship of the $U(1)$ phase to the nearby
more conventional phases.  A two-dimensional quantum Ising
version of the corner-sharing octahedra model is considered
in Appendix~\ref{app:Ising}.

\subsection{Phenomenological gauge theory description}
We want to obtain a description of the Coulomb phase
that includes the chargon fields; this is required if we
want to discuss the transition from the Coulomb phase to the 
superfluid as the charge gap collapses to zero.  
To this end, we need to consider charge carrying excitations in more 
detail.  In the large $U$ limit, this is done by deriving 
effective Hamiltonians in the charged sectors, 
just as we did in the ground state sector Eq.~(\ref{Heff}).  
For example, to study the motion of a single
chargon in a given region, we consider the sector with
$N_r = \delta_{r r_0}$ where $r_0$ can be anywhere in this region.
By inspecting the possible moves of the chargon, we find that it 
can hop only on the sites of the same sublattice of the cubic lattice.
More generally, the total chargon number on a given sublattice is 
conserved for any chargon motion that derives from the 
microscopic boson hopping.
This means that there are two distinct chargons, and can 
be also seen directly from the Gauss law Eq.~(\ref{Nr}): 
An excitation with the gauge charge $+1$ carries 
boson charge $+q_b/2$ if it resides on the $A$ sublattice and 
$-q_b/2$ if it resides on the $B$ sublattice,
and in the Coulomb phase the gauge and boson charges are both 
good quantum numbers.

Instead of working directly with the microscopically derived 
effective Hamiltonian for the chargons, which is complicated,
we consider the following model gauge theory
\begin{eqnarray}
\label{cqedw2m}
H_{2ch} &=& U \sum_r  (n_{1r}^2  +  n_{2r}^2) 
-t\sum_{rr'} (b_{1r}^\dagger b_{1r'} e^{i a_{rr'}}
+ b_{2r}^\dagger b_{2r'} e^{i a_{rr'}}) \nonumber \\
&& + u_E \sum {\bm E}^2 - K\sum \cos(\bmnabla \times \bma) ~.
\end{eqnarray}
This has two chargon fields minimally coupled to the compact $U(1)$
gauge field.  The Hilbert space of the theory is defined by
\begin{equation}
\bmnabla \cdot {\bm E} = n_{1r} + n_{2r} ~.
\end{equation} 
$b_{1r}^\dagger$ and $b_{2r}^\dagger$ both carry gauge charge $+1$, 
and we take the first to carry boson charge $+q_b/2$ and the second 
$-q_b/2$.  There is no discrimination between the two chargons, 
so the hopping amplitude and the on-site interaction are taken equal 
for the two species.
This theory has the correct chargon content and is therefore 
expected to capture the relevant physics of the microscopic model.

For later convenience, we also write down the corresponding 
(3+1)D classical action
\begin{eqnarray}
S_{2ch} &=& -\beta\sum \left[\cos(\nabla_\mu \phi_1 - a_\mu) 
                             + \cos(\nabla_\mu \phi_2 - a_\mu) \right] 
\nonumber \\
&& -K\sum \cos(\nabla_\mu a_\nu - \nabla_\nu a_\mu) ~.
\label{S2ch}
\end{eqnarray}
For simplicity, the action is written in a space-time isotropic
form, and is characterized by two coupling constants $\beta$ and $K$.
The former characterizes the relative strength of the chargon
hopping $t$ vs repulsion $U$, while the latter characterizes
the competition of $K$ vs $u_E$.

As an aside, we note that the above gauge theory also appears in a 
particular ``slave boson'' treatment of a generic bosonic Hamiltonian at 
integer filling
\begin{equation}
H_{\rm generic} = u_b \sum_r n_r^2 - w_b \sum_{rr'} b_r^\dagger b_{r'}
+ \cdots ~.
\end{equation}
Specifically, consider an analog of $CP^1$ representation for 
the bosons treated as $O(2)$ rotors
\begin{eqnarray}
b_r^\dagger \equiv e^{i\phi_r} &=& b_{1r}^\dagger b_{2r} 
\equiv e^{i(\phi_{1r}-\phi_{2r})} ~, \\
n_r &=& n_{1r} = -n_{2r}.
\label{CP1constraint}
\end{eqnarray}
Applying the corresponding slave boson mean field scheme, and studying 
fluctuations about the mean field, one obtains the above gauge theory
(the details of the derivation can be found e.g. in Appendix E of 
Ref.~\onlinecite{deccp}).  The constraint Eq.~(\ref{CP1constraint}) 
formally corresponds to the limit $u_E \to \infty$ in $H_{2ch}$, 
but we can imagine that some coarse-grained $u_E$ is finite, 
and the description in terms of $b_{1r}$ and $b_{2r}$ as emergent 
degrees of freedom applies in some regime of parameters.

\begin{figure}
\centerline{\includegraphics[width=3.5in]{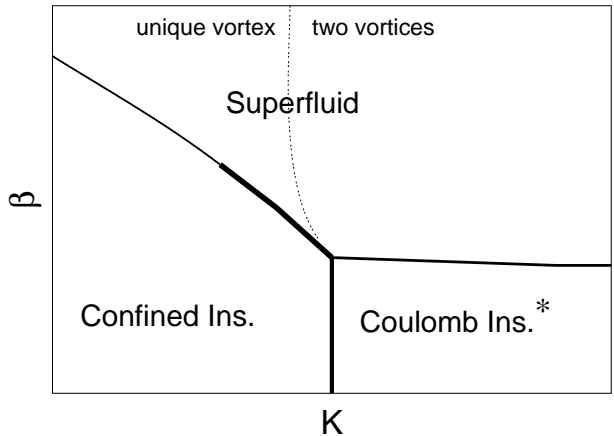}}
\vskip -2mm
\caption{Schematic phase diagram of the compact $U(1)$ gauge theory
with two matter fields, Eq.~(\ref{cqedw2m}).
The dashed line in the superfluid phase represents the conjectured
vortex core transition from a unique vortex to two physically
distinct vortices (there is no transition in the superfluid 
ground state across this line).
}
\label{cqedw2m_phased}
\end{figure}

The phase diagram of the two-chargon gauge theory $H_{2ch}$ is 
established by standard arguments\cite{FraShe} and is shown in 
Fig.~\ref{cqedw2m_phased}.
The different phases are readily identified with the phases of the 
microscopic boson Hamiltonian Fig.~\ref{phased}.
(i) In the regime $t \ll U$, $K \gg u_E$, the gauge theory is in the
deconfined Coulomb phase with gapped chargons; this is of course
the Coulomb phase of our microscopic model and has the correct
particle content by construction.
(ii) In the regime $t \ll U$, $K \ll u_E$, the gauge theory is confining
and there are no free gauge charges; only gauge-neutral excitations 
are free and carry integer boson charge;
this phase corresponds to the Mott insulator.
(iii) Finally, in the Higgs phase $t \gg U$ both chargon fields condense
and the gauge-neutral field $e^{i(\phi_1-\phi_2)}$ obtains
an expectation value; this corresponds to the superfluid phase
of the bosons.
It should be emphasized here that while the model gauge theory
reproduces the phase diagram of the physical system, it is only
in the Coulomb phase that the chargon variables represent 
low-energy degrees of freedom; in the other phases,
the chargons are confined and are not present as individual
particles in the spectrum at any energy.

The phenomenological description allows us in particular to discuss
the nature of the phase transitions.  Thus, we expect the conventional
insulator to superfluid transition to be in the (3+1)D XY universality
(bosons at integer filling) for small $K/u_E$, but it can also become 
first order for larger $K/u_E$.  The deconfinement transition between 
the two insulators is similar to that in the pure gauge theory and
empirically is believed to be first order.
Finally, in the Coulomb to superfluid transition the gauge field can be 
viewed as noncompact since the monopoles remain gapped and do not
participate.  This transition is expected to be fluctuation induced
first order.\cite{1st_order}

Our treatment so far has been entirely in the boson charge 
degrees of freedom, and to conclude this discussion
we point out some salient properties of the Coulomb phase.
First, the Coulomb phase features an emergent gauge charge
conservation law that is {\em exact} as long as the system
remains in this phase.
On a formal level, in this case, the projective transformation that 
has led us to the $U(1)$ gauge theory is convergent as a series in $t/U$,
and the result is a generic $U(1)$ gauge theory with the local
constraint that is exact by construction.
Second, the Coulomb phase and the emergent conservation law
are stable against adding {\em arbitrary} perturbations to the 
microscopic Hamiltonian as long as these perturbations are
sufficiently small;\cite{Nielsen, Wen, bosfrc, Hermele} 
this again follows from the local constraint imposed by the large $U$ 
term, since the result is by fiat a gauge theory.  
This amazing fact holds for perturbations that can break {\em all} 
symmetries of the microscopic Hamiltonian, and despite the fact that
the low-energy theory has a gapless photon.  
This is a hallmark of the topological structure present in the 
Coulomb phase.

However, the topological character of the Coulomb phase 
and its genesis appear mysterious in the charge language.
Much further insight is obtained when we consider 
vortices in the superfluid phase from the point of view of the 
phenomenological theory.

A precise topological characterization of the Coulomb phase
is provided by the existence of the gapped monopole excitation. 
At energy scales well below the monopole gap, the physics is that of a 
non-compact $U(1)$ gauge theory.  Monopoles remain gapped across the 
transition to the superfluid phase --- as we show below this leads to 
unusual vortex physics.

Specifically, we argue for the presence of two distinct stable 
vortices in the superfluid adjacent to the Coulomb phase.  
We can think of the two vortices as having distinguishable 
order in the core.  Physically, the core can change from one 
order to the other somewhere along the vortex line, 
but this domain wall costs an additional energy.
The stability of two distinct vortices in some parameter regime
is a consequence of some microscopic energetics in the system.
When this superfluid state is disordered by proliferating
both vortex species, the Coulomb phase obtains, and the domain wall 
excitations become particles identified with the gapped monopoles.

\section{Vortex phenomenology and physical pictures}
\label{sec:vortex_phenom}
The two-chargon gauge theory is amenable to a precise duality treatment, 
which gives a dynamical theory of vortices.  A formal derivation is 
given in Appendix~\ref{app:2ch}.  Here, we proceed gently and consider a 
semiclassical description of vortices in the superfluid. 
This discussion provides a physical picture that underlies the more 
formal treatments of subsequent sections. 
We argue that, depending on the microscopics, it is possible to have 
either a unique physical vortex or two distinct vortex species.
(A more careful discussion requires the full dynamical theory.)
As we vary the parameters in the superfluid phase, there is a 
transition occurring inside the core of vortex lines from a unique 
quantum core state to two degenerate but distinct states.  
The ``transition line'' cuts the superfluid phase into two regions as 
indicated in Fig.~\ref{cqedw2m_phased}.

It should be emphasized that there is no transition in the
ground state of the system.  The transition occurs in the 
properties of an infinite vortex line, which is a very high energy 
excitation.  But this has a bearing on which insulating state is obtained
when vortices condense:  If we have a unique vortex which then condenses,
we obtain the conventional insulator. 
On the other hand, as we argue below and in subsequent sections,
if we have two distinct vortex species and both condense,
we obtain the fractionalized Coulomb phase
(see Fig~\ref{cqedw2m_phased}).  
It is also possible to have two vortex species but condense only
one of them, and this will give an insulating state with broken
discrete symmetry ($b_1 \leftrightarrow b_2$ symmetry in the
two-chargon model or more physically the sublattice interchange 
symmetry in the microscopic boson model); 
in the simple lattice model Eq.~(\ref{S2ch}) this phase is 
probably not realized.

\vskip 1mm
We begin with an effective description of the superfluid phase,
which is obtained by expanding the cosines in the classical 
action Eq.~(\ref{S2ch}); the terms read schematically
\begin{eqnarray*}
(\bmnabla\phi_1 - \bma)^2 + (\bmnabla\phi_2 - \bma)^2
+ (\bmnabla \times \bma)^2 ~.
\end{eqnarray*}
Shifting $\bma \to \bma^\prime = \bma - \bmnabla\phi_2$,
the field $\bma^\prime$ is massive, and at low energies we are left 
with one Goldstone mode $[\bmnabla(\phi_1 - \phi_2)]^2$,
as expected in the superfluid phase.

To study vortices, we apply an external electromagnetic
gauge potential that couples to the physical boson number density; 
this amounts to the replacements
\begin{eqnarray*}
(\bmnabla\phi_1 - \bma - \frac{1}{2} \bmA_{ext})^2 
+ (\bmnabla\phi_2 - \bma + \frac{1}{2} \bmA_{ext})^2
+ (\bmnabla \times \bma)^2 ~.
\end{eqnarray*}
Factors $\pm 1/2$ correspond to the $\pm q_b/2$ charges carried
by the chargons.  An $hc/q_b$ vortex is obtained by requiring
\begin{equation}
\oint \bmA_{ext} \cdot d\bml = 2\pi
\end{equation}
on going around the vortex.
At large distances from the core we have
\begin{equation}
\bmnabla\phi_1 - \bma - \frac{1}{2} \bmA_{ext} = 0, \quad\quad
\bmnabla\phi_2 - \bma + \frac{1}{2} \bmA_{ext} = 0 ~.
\end{equation}
Consider configurations with $\phi_{1,2}$ winding by $2\pi m_{1,2}$
respectively, 
$\oint \bmnabla\phi_{1,2} \cdot d\bml = 2\pi m_{1,2}$.
From the above equations we conclude
\begin{eqnarray}
\oint \bmA_{ext} \cdot d\bml = 2\pi(m_1 - m_2) 
\label{m1m2}
\\
\oint \bma \cdot d\bml = \pi(m_1+m_2) ~.
\end{eqnarray}

Thus, there are different ways to realize a physical vortex in terms of 
the chargon fields.  For a unit vortex, the relevant realizations 
are $(m_1, m_2) = (1,0)$ and $(0, -1)$ since these have the lowest 
core energy.  Notice that the two realizations carry the gauge field 
flux $\Phi_a = +\pi$ and $-\pi$ respectively.
More generally, different realizations of a given vortex are 
characterized by values of the gauge flux $\Phi_a$ that differ by 
multiples of $2\pi$.
Independent of specific realizations, we note an important 
distinction between odd-strength vortices,
which carry flux $\Phi_a = \pi\, \mod\, (2\pi)$, 
and even-strength vortices, which carry $\Phi_a = 0\, \mod\, (2\pi)$.

In the above, the gauge field $\bma$ is treated as noncompact.
On a formal level, the multiplicity of the physical vortex
appears to be an artifact of ignoring compactness of the
gauge field.  Indeed, by moving a monopole across the system,
which is a dynamical degree of freedom, we can change the gauge 
field flux by $2\pi$.  For the preceding discussion of a single
vortex, configurations $(1,0)$ and $(0,1)$ that differ in 
$\Phi_a$ by $2\pi$ can mix with each other and produce a unique 
physical vortex (the mixing occurs on the level of local line
segments).  In this case, the only gauge flux distinction that remains 
is between even and odd vortices.

However, it is crucial to note that the possibility of the mixing 
discussed above is a {\em dynamical} question.  We argue below that it 
effectively does not occur near the Coulomb phase boundary.

Consider a single physical vortex.  As shown above, the vortex 
carries internal gauge flux of $+\pi$ or $-\pi$ depending on whether 
it is realized as a vortex in $\phi_1$ or $\phi_2$. 
Now consider a ``domain wall'' inside the core of such a vortex where 
the internal gauge flux changes from $+\pi$ to $-\pi$. 
As the internal gauge flux changes by $2\pi$ at such a domain wall, 
we may identify it with a monopole configuration of the gauge field. 
In the Coulomb phase this monopole costs a finite non-zero energy. 
The monopole gap is expected to stay on approaching the transition to 
the superfluid.  Consequently, close to this phase boundary but in the 
superfluid side, there will be a finite non-zero energy cost associated 
with the domain walls inside the vortex core.  This immediately implies 
the existence of two distinct vortex species with the same physical 
vorticity which are distinguished by the sign of the internal gauge flux 
in the core. 

Upon moving away from the superfluid--Coulomb phase boundary, the 
domain wall gap in the vortex core will eventually close.  The domain 
walls will then proliferate inside the core leading to the destruction 
of the vortex core order so that there will then be a unique vortex 
for any given physical vorticity. 

What is the physical meaning of the two vortex species? 
To explore this, first note that the two vortex species may be described 
as being either a full $2\pi$ vortex in $b_1$ (but not in $b_2$) or 
vice versa.  Thus the amplitude of $b_1$ is suppressed in the core of 
the $\phi_1$ vortex while that of $b_2$ is not, and the reverse happens 
for the $\phi_2$ vortex.  Thus, the order in the core of the vortex 
corresponds to spontaneously breaking the symmetry of interchange of 
$b_1$ and $b_2$.  But what is this symmetry in terms of a more 
microscopic description in terms of the underlying bosons? 
The answer clearly depends on the model considered. 
For the microscopic boson Hubbard model of Sec.~\ref{sec:real},
the two chargon species $b_{1,2}$ are associated with sites that belong 
to the two distinct sublattices of the cubic lattice on whose links the 
physical bosons live. 
Thus, the chargon interchange symmetry corresponds simply to interchange 
of the $A$ and $B$ sublattices in the underlying model.  We therefore 
conclude that this sublattice interchange symmetry (present in the 
microscopic model) is spontaneously broken in the core of a vortex line 
in the superfluid phase close to the transition to the Coulomb insulator.
The two different vortex species simply correspond to the two possible 
degenerate ordered patterns for the core state. 

An almost similar situation also obtains for bosons at half-filling 
(see Section~\ref{sec:halfint}) on a cubic lattice in a superfluid state 
close to the transition to a Coulomb insulator. 
There the two different vortex species again have broken sublattice 
symmetry in the core which may be simply identified with checkerboard 
density wave order. 
This kind of order in the vortex core is very natural for bosons at 
half-filling --- thus a Coulomb insulator is perhaps a more natural 
possibility at half-filling once the superfluidity is destroyed. 

Having understood the vortex structure in the superfluid phase, 
we can now turn around and address the question of how to view 
the Coulomb phase in terms of the excitations of the superfluid. 
To destroy the superfluidity we need to proliferate vortex lines. 
With two vortex species present, first consider proliferating just one 
kind of vortex.  The resulting insulator will then inherit the order in 
the core of the proliferated vortex species. 
In other words, the resulting insulator will break a symmetry 
(for instance, the sublattice interchange in the boson Hubbard model of 
Sec.~\ref{sec:real}). 
On the other hand, we may proliferate both vortex species with equal 
amplitudes.  The resulting insulator will then not inherit the broken 
symmetry of either vortex core.  This is the Coulomb phase. 
The monopoles are clearly the remnants of the domain walls between the 
two ordering patterns along the vortex core 
(see Fig.~\ref{fig:mnpl_roton}). 
The photon may then be associated with fluctuations of a `roton' formed 
by combining together a loop of one vortex species with a loop of an 
antivortex of the other vortex species.  To understand this, we note 
that in the gauge theoretic description of the superfluid phase such a 
roton has no net vorticity but carries internal gauge flux of $2\pi$. 
In other words, it represents a tube of magnetic flux.  Once the two 
vortex species proliferate, the fluctuations of the roton may be 
described in terms of fluctuating magnetic field lines, and this
physically describes the photon.

The above physical pictures are elaborated in precise formal terms in 
the following sections.

\section{Dual vortex description}
\label{sec:dual}
To make the above discussion of the vortex core structure and its 
implications for proximate insulating phases more precise, 
it will clearly be useful to have a dual formulation which focuses on 
the vortex line degrees of freedom. 
For the issues at the focus of the present paper --- namely a dual 
description of the Coulomb phase --- it is necessary to have a dual 
formulation that clearly brings out the presence of the two distinct 
vortex species with ordered cores.  This is most conveniently 
obtained by dualizing the gauge theoretic description in terms of the 
two chargon fields. 
Such a formulation may be obtained by an extension of the duality 
transformation for the usual lattice $XY$ model in (3+1)D.
This model describes bosons at some large integer filling on a three 
dimensional lattice. 
Since even this duality is possibly unfamiliar to most readers, 
we first review it below and show how various familiar Mott insulating 
states of bosons at integer filling may be described in terms of vortex 
line condensates.  We will then generalize this duality to incorporate 
the non-trivial core structure of the vortex lines when there are two 
vortex species, and show how the Coulomb phase results from their 
proliferation.

\subsection{Review of vortex description for (3+1)D XY model}
\label{subsec:xy}
Here we review duality for generic 3D quantum bosons at integer
filling.  The dualities are performed for the Euclidean action,
which in this case is (3+1)D XY model:
\begin{eqnarray}
S_{XY}[\phi_i] &=& -\beta \sum_{i\mu} \cos(\nabla_\mu \phi) ~.
\end{eqnarray}
To avoid clutter, we take equal spatial and temporal couplings.
The formalism is standard,\cite{Polyakov, Peskin} but we want to bring 
out its physical interpretation as was done in 
Refs.~\onlinecite{MPAFdual, Dasgupta} for (2+1)D.
Our treatment below makes explicit the correspondence of the dual 
variables to vorticities, and enables us to obtain a dual perspective 
on many familiar Mott insulating states. 
The primary purpose of this review is to provide a background
for the subsequent duality treatments of more complicated cases.

We start with the Villain form for the partition function
\begin{eqnarray}
Z_V = \sum_{[p_{i\mu}]=-\infty}^\infty \int_{-\pi}^\pi [D\phi_i]
\exp\left[-\frac{\beta}{2} \sum_{i\mu} 
          (\nabla_\mu \phi - 2\pi p_{i\mu})^2 \right] ~.
\end{eqnarray}
We work on a (3+1)D hypercubic lattice; space-time sites are labeled 
by lower-case Latin letters ($i,\dots$), and the lattice directions 
are labeled by Greek letters ($\mu, \dots$).  
Integer-valued fields $p_{i\mu}$ ensure $2\pi$ periodicity
in the angles.  For each configuration in the statistical sum,
we define vorticities
\begin{equation}
q_{\mu\nu} = \nabla_\mu p_\nu - \nabla_\nu p_\mu ~.
\end{equation}
In (3+1)D, $q_{\mu\nu}$ describe vortex worldsheets.
Here and below, the lattice derivatives are indicated only schematically,
but the precise meaning is readily recovered in each case.

As usual for $O(2)$ models, we want to separate the spin wave and 
vortex parts.  This is achieved by dividing configurations 
$\{p_{i\mu}\}$ into classes with the same vorticity.
Two configurations $\{p_{i\mu}\}$ and $\{p'_{i\mu}\}$
belong to the same class if they can be related via
$p'_{i\mu} = p_{i\mu} + \nabla_\mu N$ with an integer-valued 
field $N_i$.  Using the latter, we now extend the $\phi$ 
integrals over the whole real line and obtain
\begin{widetext}
\begin{eqnarray}
Z_V &=& {\sum_{[q_{i\mu\nu}]}}^\prime 
\int_{-\infty}^\infty [D\phi_i] \int_{-\infty}^\infty [Dj_{i\mu}]
\exp\left[-\frac{1}{2\beta} \sum_{i\mu} j_{i\mu}^2 
          +i \sum_{i\mu} j_{i\mu} (\nabla_\mu \phi - 2\pi p_{i\mu}^{(0)})
    \right] \\
&=& {\sum_{[q_{i\mu\nu}]}}^\prime \int_{-\infty}^\infty [Dj_{i\mu}]
\;\delta(\nabla_\mu j_\mu)\;
\exp\left[-\frac{1}{2\beta} \sum_{i\mu} j_{i\mu}^2 
          -i \sum_{i\mu} j_{i\mu} 2\pi p_{i\mu}^{(0)} \right] ~.
\label{xydual:ZV}
\end{eqnarray}
\end{widetext}
Here, we introduced a real-valued field $j_{i\mu}$, which can be 
interpreted as particle current;
$\{p_{i\mu}^{(0)}\}$ is any member of the class with given vorticity 
$q_{i\mu\nu}$. 
The prime on the sum over the vorticities $q_{i\mu\nu}$
indicates that these satisfy schematically 
$d (q_{\mu\nu} dx_\mu \wedge dx_\nu) = 0$,
which reads as an integer-valued constraint
\begin{equation}
\frac{1}{2} \epsilon_{\rho\sigma\mu\nu} \nabla_\sigma q_{\mu\nu} = 0
\label{novsource}
\end{equation}
for each direct lattice cube or equivalently each dual lattice link.
Here and in what follows, we specifically work in (3+1)D and make use of 
the fully antisymmetric tensor $\epsilon_{\rho\sigma\mu\nu}$ to connect
between direct and dual lattice objects; summation over repeated 
indicies is implied unless specified otherwise.
The above constraint means that there are no sources for vortices.
We can also specify the vortex worldsheets by an integer-valued
field $F_{I\rho\sigma}$ living on the dual plackets
(upper-case letters $I,\dots,$ specify dual lattice sites):
\begin{eqnarray}
F_{\rho\sigma} &=& 
\frac{1}{2}\epsilon_{\rho\sigma\mu\nu} q_{\mu\nu} ~,\\
\nabla_\sigma F_{\rho\sigma} &=& 0 ~.
\label{q2F}
\end{eqnarray}

The current conservation constraint $\nabla_\mu j_\mu=0$
can be solved by introducing a rank-2 antisymmetric field 
$\g_{I\rho\sigma}$ living on the dual plackets 
\begin{equation}
j_\mu = \frac{1}{2} \epsilon_{\mu\nu\rho\sigma} 
\nabla_\nu \frac{\g_{\rho\sigma}}{2\pi} ~,
\label{g2j}
\end{equation}
with factor $2\pi$ introduced for convenience.
We now have
\begin{equation}
\sum_{i\mu} j_{i\mu} 2\pi p_{i\mu}^{(0)} = 
\sum_{I, \rho<\sigma} \g_{I\rho\sigma} F_{I\rho\sigma}~,
\end{equation}
so vorticity appears explicitly in the statistical sum.
At this stage, we can integrate out the field $\g_{\rho\sigma}$ 
and obtain a description in terms of the vortex worldsheets with 
specific long-ranged interactions, and consider different phases of the 
vortex system as determined by this interaction.
We expect that modifying the interactions at short distances
will not change the physics of the various phases but 
only positions of the phase boundaries, and we therefore consider 
generalized models with added local ``fugacity'' terms
\begin{equation}
S_{\rm fug.} = \frac{1}{2\lambda} \sum_{i,\mu<\nu} q_{i\mu\nu}^2 
= \frac{1}{2\lambda} \sum_{I,\rho<\sigma} F_{I\rho\sigma}^2 ~.
\end{equation}

In order to describe the vortex system in more familiar terms,
we instead retain the $\g_{\rho\sigma}$ field and proceed as follows.  
The integer-valued constraint of no vorticity sources
written in terms of $F_{I\rho\sigma}$, Eq.~(\ref{q2F}), 
is handled by introducing a $U(1)$ variable $c_{I\rho}$ on 
each dual lattice link in the manner
\begin{equation}
\delta(\nabla_\sigma F_{\rho\sigma}) \propto \int_{-\pi}^\pi dc_\rho 
\exp\left[i c_\rho \nabla_\sigma F_{\rho\sigma} \right] 
\end{equation}
(no sum over $\rho$).
We can now perform the summation over vorticities obtaining
\begin{eqnarray}
Z_V[\lambda] &=& \int_{-\infty}^\infty [D{\g_{I\rho\sigma}}]^\prime
\int_{-\pi}^\pi [Dc_{I\rho}]  
\exp\left[-\frac{1}{2\beta} \sum_{i\mu} j_{i\mu}^2 \right]
\nonumber
\\
&\times & \exp\left[ \lambda \sum_{I, \rho<\sigma} 
\cos(\nabla_\rho c_\sigma - \nabla_\sigma c_\rho - \g_{I\rho\sigma})
\right] ~,
\label{xydual:final}
\end{eqnarray}
where the cosine stands for the appropriate Villain form
and $j_\mu$ is given in Eq.~(\ref{g2j}).
The only approximation in the above duality transformation
(besides Villain-izing) is the added vortex fugacity.
The resulting dual theory has a compact $U(1)$ gauge field 
coupled to a noncompact rank-2 field.  This is a generalization
of the (2+1)D duality, where the dual theory has a $U(1)$ scalar
field coupled to a noncompact gauge field.
The vorticity $q_{\mu\nu}$ of the original angle variables
is precisely the integer-valued electromagnetic field tensor 
$F_{\rho\sigma}$ of the compact electrodynamics in the $c_{I\rho}$
variables.  In a Hamiltonian formulation, vortex lines are 
identified with the electric field lines, e.g., 
$E_x \equiv F_{\tau x} \equiv q_{yz}$.
Thus, $e^{i c_{Rx}}$ is conjugate to $\hat E_{Rx}$ and
can be viewed as a vortex line segment creation operator.

For later convenience, we also exhibit the dual lattice action using
`soft-spin' vortex fields:
\begin{eqnarray}
S_{\rm vort} &=&  -\lambda \sum_{I,\rho<\sigma}
\left[\Psi_{I\rho}^* \Psi_{I+\hat\rho, \sigma}^* 
      \Psi_{I+\hat\sigma, \rho} \Psi_{I\sigma} 
\; e^{-i\g_{\rho\sigma}} + c.c. \right] \nonumber \\
&+& \sum_{I\rho} V(|\Psi_{I\rho}|^2) 
+ \kappa\sum (\epsilon_{\mu\nu\rho\sigma}\nabla_\nu \g_{\rho\sigma})^2 ~.
\label{Svort}
\end{eqnarray}
Here $\Psi_{I\rho}^* \sim e^{ic_{I\rho}}$ creates a vortex line segment 
on the dual lattice link; 
$V(|\Psi|^2) = s |\Psi|^2 + u_4 |\Psi|^4 + u_6 |\Psi|^6 + \cdots$
controls amplitude fluctuations of the field $\Psi$.
From the preceding derivation, the rank-2 field $\g_{I\rho\sigma}$ 
describes boson density fluctuations via Eq.~(\ref{g2j}).  
The first term in the action represents vortex line ``hopping'' by 
extending across a placket.  The boson density acts as a 
source for the rank-2 gauge potential seen by the vortices 
(this gauge potential produces the familiar Berry phase when a vortex 
line moves in the superfluid).

The behavior of the compact $U(1)$ gauge theory without the
rank-2 field is well understood, and we can use this as a
starting point to develop intuition about the vortex system.
This line of thinking can be interpreted as considering 
screened vortices first.  Indeed, for a charged boson system
we would conclude that the field $\g_{\rho\sigma}$ is massive 
and can be ignored.
To obtain a faithful description of neutral bosons, e.g. to correctly 
reproduce the low-energy modes, we need to include the ``gauging''
by the rank-2 field.  This can be accomplished by a semiclassical 
analysis just as in the Ginzburg-Landau theory.

We now summarize the dual description of the familiar phases
of 3D quantum bosons at integer filling.

\vskip 2mm
{\bf Superfluid} phase of the bosons is identified with
confinement in $e^{i c_\rho}$.
Vortex excitations in the superfluid are the gapped electric field
lines.  The rank-2 field $\g_{\rho\sigma}$ is essentially free, 
and its single propagating mode (which can be found e.g. by doing 
quantum mechanics for this field) is precisely the phonon mode of the 
superfluid.

\vskip 2mm
{\bf Mott Insulator} phase is obtained by proliferating vortices, 
so $e^{i c_\rho}$ is ``condensed'', or more appropriately 
``deconfined''.  The rank-2 field $\g_{\rho\sigma}$ is gapped out, 
and in the process ``eats'' the photon modes (an analog of the
Anderson-Higgs phenomenon), so there are no gapless excitations, 
as expected.  Gapped charged excitations of the Mott insulator are
represented by monopoles of the dual gauge field,
and the discreteness of charge is encoded in the 
corresponding quantization condition for the rank-2 field,
which is a generalization of the flux quantization in the 
Ginzburg-Landau theory.

Indeed, let us consider a monopole worldline.  Fix a time slice
(of our Euclidean path integral) and evaluate the boson number
\begin{equation}
n_{\rm tot} = \int j_{\tau} d^3 r = \frac{1}{2\pi} \int d^3 r
(\nabla_x \g_{yz} + \nabla_y \g_{zx} + \nabla_z \g_{xy}) ~. 
\end{equation}
In the far field away from the monopole we have e.g.
\begin{equation}
\g_{xy} = \nabla_x c_y - \nabla_y c_x \equiv B_z ~,
\end{equation}
so the right hand side coincides with the outgoing magnetic
flux divided by $2\pi$, which is equal to the enclosed monopole 
number.  Thus, monopole worldlines in the dual description
are precisely worldlines of integer-quantized charges in the 
direct description.

\vskip 2mm
{\bf $Z_2$ Fractionalized Insulator}:
We can readily generalize the above description of the Mott insulator
to discuss more exotic insulator phases.  
$Z_2$ fractionalized phase is obtained as a condensate of doubled
vortices while single vortices are not condensed.  
The dual action generically allows double-vortex hopping 
terms like
\begin{equation}
-\lambda_2  \sum_{I,\rho<\sigma} 
\cos\left[ 2 (\nabla_\rho c_\sigma - \nabla_\sigma c_\rho 
              - \g_{I\rho\sigma}) \right] ~.
\label{lambda2}
\end{equation}
When $\lambda$ is small while $\lambda_2$ is large,
we expect $e^{i 2 c_\rho}$ to be deconfined but not $e^{i c_\rho}$.
Topological excitations in this phase are monopoles of the  
field $e^{i 2 c_\rho}$, and these are seen to represent
charged excitations carrying fraction $1/2$ of the boson charge.
Odd-strength vortex line excitations correspond to vison line
excitations of the $Z_2$ phase.


\subsection{Dual formulation with two vortex species}
\label{subsec:2vort}
We now show how to incorporate the physics of two distinct vortex 
species with ordered cores in the dual formulation.  As mentioned 
above, this is most conveniently done by dualizing the gauge theoretic 
description in terms of two chargon fields.  The duality transformation 
is a straightforward but tedious extension of the methods of the 
previous subsection.  We therefore relegate the technical details to 
Appendix~\ref{app:2ch} and focus here on the result and the physics 
contained in it.
As expected on the basis of the semiclassical discussion in 
Sec.~\ref{sec:vortex_phenom}, the analysis in Appendix~\ref{app:2ch} 
yields a dual theory with two vortex fields $\Psi^{(1)}$ and 
$\Psi^{(2)}$ corresponding to the two possible ordering patterns in the 
core.  Further, these vortices are allowed to turn from one flavor into 
the other, and the corresponding domain wall appears as an additional 
field $\Upsilon$ in the dual theory. 

A generic `soft-spin' action for the two vortex fields can be expressed
in the form
\begin{eqnarray}
S_{\rm 2vort}
&=& -\lambda \sum_{b=1,2} 
\left[\Psi^{(b)*}_{I\rho} \Psi^{(b)*}_{I+\hat\rho, \sigma}
      \Psi^{(b)}_{I+\hat\sigma, \rho} \Psi^{(b)}_{I\sigma} 
\; e^{-i \g_{\rho\sigma}} + c.c. \right] \nonumber \\
&+& \sum \left[ V(|\Psi^{(1)}_{I\rho}|^2 + |\Psi^{(2)}_{I\rho}|^2)
               + W(|\Psi^{(1)}_{I\rho}|^2 |\Psi^{(2)}_{I\rho}|^2)
         \right]
\nonumber \\
&+& 
\kappa \sum (\epsilon_{\mu\nu\rho\sigma}\nabla_\nu \g_{\rho\sigma})^2 
\nonumber \\
&-& \lambda_m \sum
(\Upsilon_I^* \Upsilon_{I+\hat\rho} 
 \Psi^{(1)*}_{I\rho}\Psi^{(2)}_{I\rho} 
 + c.c.) \nonumber \\
&+& \sum U(|\Upsilon_I|^2) ~.
\label{S2vort}
\end{eqnarray}
In the two-chargon theory, both $\Psi^{(1)*}_{I\rho}$ and 
$\Psi^{(2)*}_{I\rho}$ create a physical vortex line segment, 
but it is realized as a vortex in $b_1$ in the first case and 
an antivortex in $b_2$ in the second case, see Eq.~(\ref{m1m2}).
The vortex interaction terms $V$ and $W$ need not be specified at this 
stage but are important when determining the vortex condensate.
$\lambda_m$ represents quantum tunneling between the two core states
and is written here as the domain wall hopping, while $U$ represents 
the domain wall energy cost.

Two broad possibilities need to be considered in the superfluid.
The first possibility is that domain walls are energetically 
cheap and proliferate inside vortex cores.  
In this case the two vortex fields lock to each other.  
In the integer filling case considered here, this produces a unique 
physical vortex, and the proper low-energy description is then in terms 
of a single vortex field
(the situation at half-integer filling is more complicated, 
see the next section).

The second possibility is that domain walls are energetically
costly and remain gapped inside vortex cores.  In this case 
the field $\Upsilon$ has only short ranged correlations and may be 
integrated out.  The two vortex fields are proper low-energy degrees of 
freedom.  We can now ask what happens when these vortices condense, 
e.g. when $\Psi=0$ becomes an unstable state of the potential $V$.
Depending on the interaction $W$, the condensate either has just 
one of the two vortex species, or has both vortex species proliferated
simultaneously.  
The former case occurs when $W>0$, and we write schematically 
$\la \Psi^{(1)} \ra \neq 0$, $\la \Psi^{(2)} \ra = 0$.  
The resulting Mott insulator then inherits the order in the core of 
the $\Psi^{(1)}$ vortex.   There are no gapless modes: as discussed in 
the previous subsection, the rank-2 field Higgses out the gapless modes 
of the single-species worldsheet condensate.
In the microscopic boson Hubbard model on the lattice of corner-sharing 
octahedra, this Mott insulator breaks $A \leftrightarrow B$ sublattice 
interchange symmetry.

On the other hand, for the interaction $W$ of the opposite sign,
it is favorable for both vortex species to proliferate simultaneously, 
schematically $\la \Psi^{(1)} \ra = \la \Psi^{(2)} \ra \neq 0$.
In this case we obtain fractionalized Coulomb insulator with a 
gapless photon, a gapped monopole, and two species of gapped chargons.  
The gapless photon obtains since the rank-2 field $\g_{\rho\sigma}$ 
can Higgs out only the $\Psi^{(1)*} \Psi^{(2)*}$ `part' of the 
two-species worldsheet condensate.  The $\Psi^{(1)*} \Psi^{(2)}$ 
part does not couple to $\g_{\rho\sigma}$; it represents string objects 
with short-range interactions only --- indeed, it is precisely the roton 
formed by combining a vortex of one kind with an antivortex of the other 
kind.  To show that its fluctuations lead to a gapless photon mode, 
we focus on just the fluctuations of the `phases' 
$\Cscr^{(1)}, \Cscr^{(2)}$ of the two vortex species:
\begin{equation}
\Psi^{(1)*} \sim e^{i\Cscr^{(1)}}, \quad
\Psi^{(2)*} \sim e^{i\Cscr^{(2)}} ~.
\end{equation}
[Note that compared with Appendix~\ref{app:2ch}, we use
$\Cscr^{(1)} = c^{(1)}$, $\Cscr^{(2)} = -c^{(2)}$, while the
monopole field $\theta^{(m)}$ is the `phase' of the domain wall
particle $\Upsilon_I^* \sim e^{i\theta^{(m)}_I}$.]
Expanding to quadratic order in small gradients of 
$\Cscr^{(1)}, \Cscr^{(2)}$, the resulting `elastic' action 
has the following schematic structure:
\begin{eqnarray}
S & = & (\bmnabla \times \Cscr^{(1)} - \g)^2
      + (\bmnabla \times \Cscr^{(2)} - \g)^2 \\
&+& \left[\bmnabla \times (\Cscr^{(1)} - \Cscr^{(2)})
          \right]^2 + \dots 
\end{eqnarray}
Here $\g$ represents the rank-2 gauge field.  The ellipses refer to 
the kinetic energy term for the $\g$ and various anharmonic corrections 
to the above quadratic action.  It is now easy to see that the 
combination $\Cscr^{(1)} + \Cscr^{(2)}$ is rendered massive while 
the combination $\Cscr^{(1)} - \Cscr^{(2)}$ remains massless. 
The latter precisely represents a linear dispersing gapless photon with 
two transverse polarizations in three spatial dimensions. 
Thus as promised the photon is indeed associated with fluctuations of 
the roton formed from a vortex of one kind and antivortex of the other 
kind.  The gapped $\Upsilon$ field simply corresponds to the monopole
excitation of the Coulomb phase.  
Finally, the ``monopoles'' of the proliferated dual fields 
$\Psi^{(1)}$ and $\Psi^{(2)}$ are the two chargons.  This completes our 
formal description.
We also point that Appendix~\ref{app:Ising} develops a similar
description of the Ising $P^*$ phase in two spatial dimensions that
incorporates the physics of two distinct Ising domain walls.

We conclude by emphasizing that the vortex proliferation transition is 
likely first order, in which case the lattice action for vortices
does not have a formal continuum limit, but this does not affect the 
presented physical picture of the phases.

\section{Bosons at half-filling}
\label{sec:halfint}
We now consider bosons at half-filling in some detail.  
We have in mind some generic Hamiltonian
\begin{equation}
H_{\rm generic} = u_b \sum_r \left(n_r-\frac{1}{2}\right)^2 
- w_b \sum_{rr'} b_r^\dagger b_{r'} + \cdots ~.
\end{equation}
The physical picture presented in the Introduction and elaborated
in the previous section carries over readily and is perhaps more 
appealing in this case.
In particular, the ordering pattern in the vortex core when there are 
two species simply corresponds to checkerboard density order of the 
bosons.  This is a very natural ``insulating'' core for the vortices 
to develop when the bosons are at half-filling. 
Formally, this core structure may readily be seen to arise in the same
semiclassical treatment as in Sec.~\ref{sec:vortex_phenom}. 
We first break up the boson field into two `chargon' fields as before. 
This leads to a description in terms of a compact $U(1)$ gauge theory 
coupled to two chargon fields.  A physical vortex in the superfluid 
order then corresponds to having a full vortex in $b_1$ (but none in 
$b_2$) or a full antivortex in $b_2$ (and no vorticity in $b_1$). 
Thus, there are still two types of vortices which convert into each 
other at the locations of monopoles of the gauge theory. 
Following the closely related discussion in Ref.~\onlinecite{deccp} in 
two dimensions, it is readily seen that the two vortices correspond to 
the two possible patterns of checkerboard density wave ordering in the 
core.  (The precise reason is simply that the two vortices differ in the 
sign of the internal gauge flux.  For bosons at half-filling, 
this gauge flux couples linearly to the difference of the boson 
densities on the two sublattices; hence there is checkerboard density 
wave ordering in the core). 

As for bosons at integer filling, the insight into the vortex structure 
can be given a precise form by employing a duality transformation.
The dual action is derived in Appendix~\ref{app:2ch_halfint}.
The Coulomb phase is again obtained when both vortex species condense 
with equal amplitude. 
The domain walls in the ordered vortex core survive as gapped monopoles, 
and the photon emerges exactly as before (as the fluctuations of 
proliferated rotons made by combining a vortex and an antivortex of two 
different species).  The gapped chargons are again ``monopoles'' of the 
two vortex fields that appear in the dual theory. 
Thus the main structure of the description of the Coulomb phase is 
unaffected from that appropriate at integer filling. 

The primary differences arise when we consider the nature of the 
possible confined phases.  These may be obtained from the Coulomb phase 
by condensing the gapped monopoles. 
Such a condensation leads to confinement of the gapped chargons and the 
disappearance of the photon from the spectrum.  However, as can be 
anticipated from experience in two dimensions, the monopoles now 
transform non-trivially under lattice symmetries.  This leads to broken 
lattice symmetry in the confined phases.  In particular, various bond or 
box density wave phases are possible.  These will be explored 
directly in the next section.

Here we specify more precisely the modifications to the two-vortex 
action Eq.~(\ref{S2vort}) due to half-filling on a cubic lattice and 
sketch some implications for the vortex core structure.
As discussed in Appendix~\ref{app:2ch_halfint}, there is an additional 
static vector potential $X^0_{I\rho}$ that couples to the domain wall
hopping term --- the $\lambda_m$ term in Eq.~(\ref{S2vort}) --- see
Eq.~(\ref{includeX0}).  
This vector potential encodes the monopole Berry phases due to 
half-integer boson density.  It is specified by the corresponding fluxes 
$\h^0_{\rho\sigma} = \nabla_\rho X^0_\sigma - \nabla_\sigma X^0_\rho$
(modulo $2\pi$) through the faces of the dual cubes.  
The fluxes are nonzero on the spatial plackets only, and equal to 
$\pi/3$ when oriented from the $A$ to $B$ sublattice of the direct 
cubic lattice as shown in Fig.~\ref{h0gauge}.
The derivation of this result is given in the next section, 
and further details can be found in Appendix~\ref{app:2ch_halfint}.

As discussed above, the two fields $\Psi^{(1)}$ and $\Psi^{(2)}$
correspond to the two checkerboard charge orders inside vortex core.
We can now ask about the effect on the core dynamics due to the static
vector potential $X^0_\rho$ seen by the domain walls.
The specific details of the domain wall motion such as change in the 
dispersion are not immediately relevant for the low-energy physics if 
the domain wall remains gapped.  Indeed this is why the modifications 
induced by half-filling do not affect much the description of the 
Coulomb phase itself.  But they become more important when 
the monopole gap closes at a phase transition from the Coulomb to a 
confined phase, or when the domain wall gap closes inside vortex core.

Deep in the superfluid phase when the vortex line is stiff, we expect
a single minimum in the domain wall dispersion, and when the walls proliferate 
inside the core we obtain a unique vortex.
However, for a vortex line that fluctuates sufficiently and
explores the three-dimensional space so that the effects of $X^0_\rho$ 
are felt, the bottom of the domain wall band can split.
This is the situation in the Coulomb phase, where domain walls 
become monopole particles, and one finds two low-energy propagating 
monopole modes (see the next section for details).

Consider now what happens when we have two domain wall band minima 
inside the vortex core in the superfluid phase, and the domain 
wall gap goes to zero.
One possibility motivated by our analysis in the next section is 
that domain wall proliferation leads to a VBS order inside the core.  
In this case, while the fields $\Psi^{(1)}$ and $\Psi^{(2)}$ are 
locked, there can be several degenerate ways for this to happen 
corresponding to the number of favorable VBS states inside the core
(we need to be even more careful here since e.g. for the columnar order 
the favorable states likely depend on the vortex line orientation).  
This would mean that our choice of the basic fields $\Psi^{(1)}$ and 
$\Psi^{(2)}$ is not a very good one, and the analysis needs to be 
reconsidered, perhaps introducing different vortex fields.
When a single such vortex condenses, the corresponding VBS state
results, but it may be also possible to condense several such vortices
simultaneously possibly leading to new exotic states.  
Even more exotic possibility is for the domain walls to become critical 
inside the core, which might occur due to the one-dimensionality of 
the setting.\cite{halfint_direct}
In the present work, we have only touched upon these possibilities 
and have not pursued any systematic studies.

\vskip 1mm
We now comment on the possibility of a direct analysis of a 
more `microscopic' vortex theory for bosons at half-filling and the 
connection with the above two-vortex description.
Indeed such a direct analysis is very useful in two spatial 
dimensions.\cite{Lannert, deccp}  The bare vortex action at half-filling
is described in Appendix~\ref{app:halfint}.  As in (2+1)D,
the half-filling manifests itself by ``frustrating'' the vortex
propagation with additional static Berry phases.  In (3+1)D, this is encoded 
by placing half of a magnetic monopole inside each dual cube in the 
compact gauge field part that describes the vortex motion.
As mentioned in Appendix~\ref{app:halfint}, at present we do not know 
how to analytically treat such frustrated gauge theories and how to 
connect directly with the two-vortex description like it was done in 
Ref.~\onlinecite{Lannert} for (2+1)D.  
Our earlier discussion and the analysis in Appendix~\ref{app:2ch_halfint}
provide an alternative route in two dimensions\cite{deccp} that extends 
reasonably to three dimensions, and we expect the resulting action 
with two vortex species to describe the vortex physics at low energies 
in some regime of parameters.
(A limited but direct attack is possible on the bare vortex action 
as well.\cite{halfint_direct})

\section{Valence bond solids of bosons at half-filling
on a 3D cubic lattice}
\label{sec:vbs}
In this section, we show how valence bond solid phases emerge due to 
monopole condensation out of the Coulomb phase at half-filling.  
The crucial ingredient is the monopole dynamics produced by the 
static vector potential $X^0_\rho$.
To that end it is convenient to specialize to the hard-core limit in 
which case we may view the bosons as representing spin-$1/2$ moments
living on the cubic lattice.  In this limit, the static vector potential 
may be viewed as arising from Berry phases present in a path integral 
description of a spin-1/2 moment.  Because of the importance of this 
result, which is an extension to three dimensions of Haldane's 
calculation in two dimensions,\cite{Haldane} 
we give a direct derivation in the context of spin-1/2 systems.
It is in this context that the valence bond solids were analyzed in the 
work of Read and Sachdev,\cite{RSvbs} providing us with familiar
grounds.

\subsection{Monopole action and Berry phases}
Our starting point is the `modified' Sachdev-Jalabert\cite{SJ, sp, deccp}
lattice model for spin-1/2 system on a cubic lattice in the vicinity
of the collinear Neel state.  This is defined in terms of a bosonic 
field $z_{i\alpha}$ which carries spin-$1/2$ and lives on the 
sites (denoted  $i,j,.....$) of a (3+1)D space-time lattice 
($\alpha = \uparrow, \downarrow$ is a spin index). 
We will refer to the quanta of this field as spinons. 
They are the analog (for the spin system) of the chargon fields 
introduced earlier for boson models.  The spinons are minimally coupled 
to a compact $U(1)$ gauge field $a_{i\mu}$. 
The action for the Sachdev-Jalabert model reads 
\begin{eqnarray*}
Z_{SJ} &=& \sum_{[u_{i\mu\nu}]} \int_{-\pi}^\pi [Da_{i\mu}]
\int [D{\bf z}_i] [\delta(|{\bf z}_i|^2-1)] \\
&&~~~~~~~ \times \exp\left(-S_z - S_a - S_B\right) ~,
\end{eqnarray*}
\begin{eqnarray*}
S_z &=& -\beta \sum_{i\mu}
[{\bf z}_i^\dagger  e^{ia_{i\mu}} {\bf z}_{i+\hat\mu} + c.c.]
+ \cdots ~,\\
S_a &=& \frac{K}{2} \sum_{i, \mu<\nu}
\left(\nabla_\mu a_\nu - \nabla_\nu a_\mu - 2\pi u_{i\mu\nu}\right)^2 
~,\\
S_B &=& i \sum_i \eta_i a_{i\tau} ~.
\end{eqnarray*}
The compactness of the gauge field is encapsulated by summing over 
integer-valued fields $u_{i\mu\nu}$.  The last term represents a Berry 
phase that encapsulates the spin-$1/2$ nature of the moment at each site.
The $\eta_i$ appearing in this term is $+1$ on the $A$ sublattice and 
$-1$ on the $B$ sublattice of the spatial cubic lattice. 
Connection to microscopic spin models and motivation may be found in the 
original references~\onlinecite{RSvbs, SJ, sp}.  With appropriate 
deformations of $S_z$, this model can also describe bosonic systems at 
half-integer filling (see Refs.~\onlinecite{sp, deccp}).  

Here we are interested in paramagnetic states with gapped spinons
(interpreted as chargons in bosonic systems).  In such cases we may 
integrate out the spinon fields completely.  This effectively amounts to 
dropping the $S_z$ term from the action.  (As we are interested in the 
general structure of the paramagnetic phases, we will not worry much 
about effects such as generating other short ranged interactions for the 
gauge fields).  The remaining action is that of a compact $U(1)$ gauge 
theory in (3+1)D on a cubic lattice but in the presence of the Berry 
phase term.  The latter has a simple physical interpretation in a 
Hamiltonian language.  It simply corresponds to the presence of a static 
background gauge charge of strength $\pm 1$ on the sites of the spatial 
cubic lattice.  The sign alternates from one sublattice to the other. 
Indeed precisely the same kind of gauge theory arises in studies of 
two-dimensional spin-$1/2$ quantum paramagnets. 

In the (2+1)D case, it is well-known that the alternating static 
background charge leads to broken translational symmetry in confined 
paramagnetic phases.  
Our goal is to extend that analysis to three dimensions. 
To that end we perform a duality transformation on $S_a + S_B$.
In the absence of the background charge, this is just the familiar 
electric-magnetic duality. 
The compact $U(1)$ gauge theory then becomes a theory of point-like 
monopoles that are coupled to the dual fluctuating non-compact $U(1)$ 
gauge field.  (The non-compactness of the dual gauge field is because 
we have thrown out the spinons --- which are the gauge `electric' 
charges --- in the original theory.  Indeed retaining them is equivalent
to retaining monopoles in the dual gauge field).  The presence of the 
background charge means that there is dual magnetic flux emanating out 
of the center of each cube on the dual lattice on which the monopoles 
live.  This flux alternates in sign from one cube to the next and 
frustrates the monopole hopping.  We therefore expect the duality 
transformation on $S_a + S_B$ to yield a theory of monopoles with 
frustrated hopping that is coupled to the dual non-compact gauge field.
This is explicitly shown in what follows. 
 
We proceed by writing
\begin{eqnarray*}
e^{-S_a} &=& \int_{-\infty}^\infty [Df_{i\mu\nu}] 
e^{-\frac{1}{2K} \sum_{i, \mu<\nu} f_{i\mu\nu}^2 } \\
& \times & 
\exp[ i\sum_{i, \mu<\nu} f_{i\mu\nu} 
      (\nabla_\mu a_\nu - \nabla_\nu a_\mu - 2\pi u_{i\mu\nu}) ] ~.
\end{eqnarray*}
$f_{\mu\nu}$ can be interpreted as the electromagnetic field tensor.
We classify configurations $\{ u_{i\mu\nu} \}$ by their monopole 
four-currents
\begin{equation}
J^{(m)}_{\rho} = \frac{1}{2} \epsilon_{\rho\sigma\mu\nu} 
\nabla_\sigma u_{\mu\nu} ~.
\end{equation}
Two configurations $\{ u_{\mu\nu} \}$ and $\{ u_{\mu\nu}^\prime \}$
belong to the same class if they can be related
$u_{\mu\nu}^\prime = u_{\mu\nu} + \nabla_\mu V_\nu - \nabla_\nu V_\mu$
with an integer-valued field $V_\mu$.  This allows separation of the 
Gaussian and topological defect parts of the statistical sum, 
since the $a_\mu$ integration now extends over the full real line.

To treat the Berry phase term, we find a static $f^0_{\mu\nu}$ 
satisfying
\begin{eqnarray}
\eta_i \delta_{\mu\tau} &=& \nabla_\nu f^0_{\mu\nu} ~, 
\label{f0:1} \\
f^0_{\mu\nu} &=& -\frac{n_0 K}{\beta}
(\delta_{\mu\tau} \nabla_\nu \eta_i 
 - \delta_{\nu\tau} \nabla_\mu \eta_i) ~.
\label{f0:2}
\end{eqnarray}
(As will be clear below, the proportionality constant $n_0 K/\beta$ 
is in fact fixed by the above conditions to be $1/12$; 
the specific writing is chosen for convenience when reusing the 
present analysis in Appendix~\ref{app:2ch_halfint}.)
The first condition is possible since 
$v_{i\mu} \equiv \eta_i \delta_{\mu\tau}$ has zero divergence.
It allows us to write the Berry phase term as
\begin{equation}
S_B = i \sum_{\mu<\nu} f^0_{\mu\nu} 
(\nabla_\mu a_\nu - \nabla_\nu a_\mu) ~.
\end{equation}
We can now bring out the topological defect part by changing 
variables to $\tilde{f}_{\mu\nu} = f_{\mu\nu} - f^0_{\mu\nu}$.
In terms of the variables $\tilde{f}_{\mu\nu}$ the action has the 
same form as in the absence of the original Berry phases except for 
additional contributions
\begin{equation}
S_{n_0} + \tilde{S}_B \equiv
\frac{1}{K} \sum_{\mu<\nu} \tilde{f}_{\mu\nu} f^0_{\mu\nu}
+ i \sum_{\mu<\nu} f^0_{\mu\nu} 2\pi u_{\mu\nu} ~.
\end{equation}
Due to condition~(\ref{f0:1}) the 'phase' term $\tilde{S}_B$ depends 
on the monopole configuration only.  The second condition~(\ref{f0:2}) 
allows us to write the term $S_{n_0}$ as
\begin{equation}
S_{n_0} = \sum_{i} \frac{n_0}{\beta} \eta_i j_{i\tau} ~.
\label{Sn0}
\end{equation}
Here $j_\mu \equiv \nabla_\nu \tilde{f}_{\mu\nu}$ can be viewed
as the gauge charge current.  This follows by integrating over the 
$a_\mu$ field;  Appendices~\ref{app:2ch}~and~\ref{app:2ch_halfint} 
exhibit this explicitly for $U(1)$ matter fields, while in the absence 
of any matter we get $\nabla_\nu \tilde{f}_{\mu\nu} = 0$.
Thus, $S_{n_0}$ can be viewed as a static staggered chemical 
potential seen by the ${\bf z}$ fields, and as long as it is much 
smaller than the spinon gap, it can be ignored for the low-energy 
properties.

We now exhibit $f^0_{\mu\nu}$ that solves the two conditions
Eqs.~(\ref{f0:1})~and~(\ref{f0:2}).  Only the electric field parts 
$E^0_x \equiv f^0_{\tau x}$, $E^0_y$, and $E^0_z$ are nonzero and
satisfy
\begin{eqnarray}
\bmnabla \cdot {\bm E}^0 &=& \eta_r ~,\\
{\bm E}^0 &=& - \frac{K n_0}{\beta} \bmnabla \eta_r ~. 
\end{eqnarray}
This fixes ${\bm E}^0$ uniquely to have value $1/6$ on each
bond from the $A$ to $B$ sublattice 
(the value $n_0/\beta$ is also fixed at this stage).

\begin{figure}
\centerline{\includegraphics[width=3.0in]{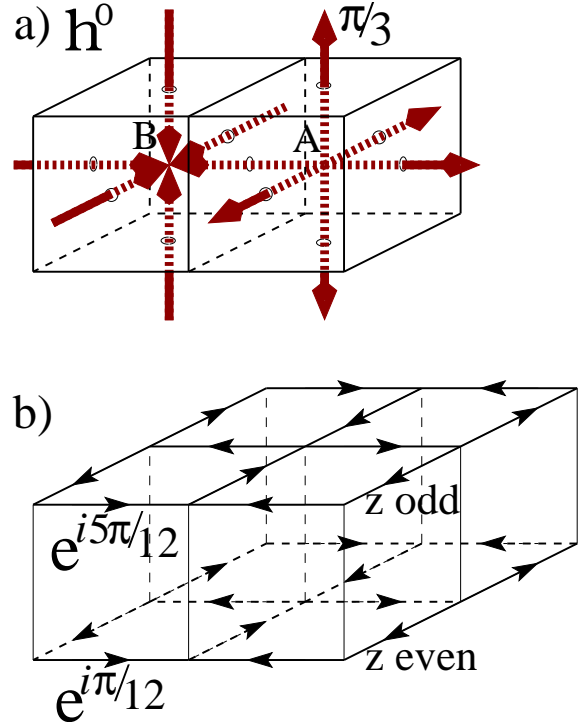}}
\vskip -1mm
\caption{
Description of the monopole Berry phases for bosons at half-filling
on a simple cubic lattice in terms of a static gauge potential
seen by the monopoles hopping on the dual lattice.
a) $\h^0_{\rho\sigma}$ gives a flux of $\pm \pi/3$ through each 
spatial placket of the dual lattice with the flux oriented
from the $A$ to $B$ sublattice of the direct lattice.
b) Our gauge choice for $e^{iX^0_{\rho}}$ that realizes
the fluxes $e^{i \h^0_{\rho\sigma}}$ (Eq.~\ref{X0}).
}
\label{h0gauge}
\end{figure}

The phase term $\tilde{S}_B$ can be expressed in terms of the 
monopole currents $J^{(m)}_\rho$ by first transforming from 
$f^0_{\mu\nu}$ to $\h^0_{\rho\sigma}$:
\begin{equation}
f^0_{\mu\nu} = \frac{1}{2} \epsilon_{\mu\nu\rho\sigma}
\frac{\h^0_{\rho\sigma}}{2\pi} ~.
\end{equation}
The latter has spatial components $\h^0_{xy}, \h^0_{yz}, \h^0_{zx}$
only, which can be interpreted as fluxes through the faces of the
dual lattice cubes.  The total outwards flux is $+2\pi$ for a dual cube 
surrounding an $A$-sublattice site of the direct lattice and 
$-2\pi$ surrounding a $B$-sublattice site, and in each case the
flux is divided equally among all six faces.  
This is shown in Fig.~\ref{h0gauge}a.
Since the total flux out of each cube is $\pm 2\pi$, we can write
\begin{equation}
\label{X0}
\h^0_{\rho\sigma} = \nabla_\rho X^0_\sigma - \nabla_\sigma X^0_\rho 
\quad\mod\quad 2\pi ~.
\end{equation}
$X^0_\rho$ is a static field on the spatial links of the 
dual lattice, and one choice is shown in Fig.~\ref{h0gauge}b.
We finally obtain
\begin{equation}
\tilde{S}_B = i \sum J^{(m)}_\rho X^0_\rho
\end{equation}
and interpret this as a monopole Berry phase.
This is appropriate since as we have explained the remaining
contributions to the low-energy action have the same form as in the 
absence of the original Berry phase $S_B$.

We can offer the following geometric representation of the monopole
Berry phase.  Consider a monopole space-time worldline and trace out 
the spatial path traveled by the monopole.  The Berry phase is given
by the flux of $\h^0$ (see Fig.~\ref{h0gauge}) through any surface pulled
onto this closed path.  This is (3+1)D generalization of Haldane's
(2+1)D result\cite{Haldane} for the monopole Berry phase.
In (2+1)D, the Haldane's result leads to destructive interference
among monopole worldpaths unless monopoles are quadrupled.
In (3+1)D, the interference pattern encoded in $\h^0_{\rho\sigma}$ 
is more subtle, and requires us to study the monopole hopping in
the static vector potential $X^0_\rho$; this is detailed below.

We now complete the duality mapping for the model $S_a+S_B$ without
any matter field in order to exhibit the relevant structure of
the monopole action; the relationship of this to the original
Sachdev-Jalabert model will be commented upon later.

Using the standard result for the monopole action in compact QED 
we obtain
\begin{eqnarray*}
Z_{{\rm cqed}+S_B} &=& {\sum_{[J^{(m)}_\rho]}}^\prime 
\int_{-\infty}^\infty [DL_\rho]^\prime 
e^{-\frac{1}{8 K\pi^2} \sum_{\rho<\sigma} 
   (\nabla_\rho L_\sigma - \nabla_\sigma L_\rho)^2} \\
&& \times e^{-i \sum J^{(m)}_\rho (L_\rho + X^0_\rho)} ~. 
\end{eqnarray*}
Here, $L_\rho$ is a noncompact dual gauge field such that
$\tilde{f}_{\mu\nu} = \epsilon_{\mu\nu\rho\sigma} 
\nabla_\rho L_\sigma/(2\pi)$.
The prime on the sum indicates that the monopole currents
satisfy continuity $\nabla_\rho J^{(m)}_\rho = 0$.
Integrating over $L_\rho$ we would obtain an action in terms of the 
monopole worldlines only.
Instead of doing this, we transform the monopole action into
a more familiar form by considering a generalized model
with added monopole fugacity term
\begin{equation}
S_{\rm fug.} =  \frac{1}{2\lambda_m} \sum (J^{(m)}_\rho)^2 ~.
\end{equation}
The constraint $\nabla_\rho J^{(m)}_\rho (I) = 0$ on each dual site $I$ 
can be solved by introducing a $U(1)$ field $\theta^{(m)}_I$, 
and upon summing over $J^{(m)}_\rho$ the final result reads
\begin{equation}
Z_{{\rm cqed}+S_B}[\lambda_m]
= \int_{-\infty}^\infty [DL_{I\rho}]^\prime 
  \int_{-\pi}^{\pi} [D\theta^{(m)}_I] e^{-S[L_\rho, \theta^{(m)}]} ~,
\end{equation}
\begin{eqnarray}
S[L_\rho, \theta^{(m)}] &=& \frac{1}{8 K\pi^2} 
\sum_{\rho<\sigma} (\nabla_\rho L_\sigma - \nabla_\sigma L_\rho)^2 
\nonumber \\
&-& \lambda_m \sum \cos(\nabla_\rho \theta^{(m)} - X^0_\rho - L_\rho) ~.
\label{Smnpl}
\end{eqnarray}
$e^{i\theta^{(m)}}$ can be interpreted as a monopole creation
operator; the field $\theta^{(m)}$ is coupled to the noncompact 
(dual) gauge field $L_\rho$, and there is also static 
``frustration'' $X^0_\rho$ coming from the Berry phases for 
monopoles in our spin-1/2 system.

The relationship of the above action to the original Sachdev-Jalabert 
model is as follows.  The action was derived in the absence of the
spinon field ${\bf z}$ (e.g. taking the spinon gap to infinity).
Now, ${\bf z}$ carries electric charge of the original gauge field 
$a_\mu$, and as usual in the electromagnetic duality, it acts as a 
magnetic charge for the dual gauge field $L_\rho$.  
Thus, to have a complete correspondence with the original spin model, 
we need to allow monopoles in the dual gauge field $L_\rho$,
and these monopoles need to be spin-1/2 particles.
Ignoring the spinon field makes the dual gauge field noncompact as 
in the above action.

\subsection{Monopole condensation patterns}
We now study possible phases of the above monopole action
assuming throughout that the spinons are gapped.

Clearly, for small $\lambda_m$ the monopole field is gapped.
In this case, the dual gauge field is free, and we obtain a 
$U(1)$ spin liquid phase of the original spin-1/2 problem.
This phase has gapped spinons, gapped monopoles, and 
a gapless photon, and respects all lattice symmetries.

To analyze possible monopole condensates, we closely 
follow Ref.~\onlinecite{Lannert}.  We first ignore
fluctuations of the dual gauge field $L_\rho$
and study a frustrated $XY$ model defined by the static 
vector potential $X^0_\rho$.  What we are essentially
doing here is averaging over fast fluctuations imposed
by $X^0_\rho$.  Once the structure of the resulting
slow fluctuation description becomes clear, 
we will restore the field $L_\rho$.

Continuous-time soft-spin action for the frustrated XY model is
\begin{equation}
\int d\tau \Big[
\sum_R |\partial_\tau \Phi_R|^2
-\sum_{\la RR' \ra} (t_{RR'} \Phi_R^* \Phi_{R'} + c.c.)
+\sum_R V(|\Phi_R|^2) \Big] ~,
\end{equation}
with some potential 
$V(|\Phi_R|^2) = r_0 |\Phi_R|^2 + u_0 |\Phi_R|^4 + \cdots.$
We work on the discrete spatial lattice labeled by $R$ 
(dual to the original spin system lattice),
since the lattice is crucial at this stage.

The frustration is encoded in the monopole hopping amplitudes 
$t_{RR} = t e^{i X^0_{RR'}}$, corresponding to the fluxes 
$\h^0$ on the plackets as shown in Fig.~\ref{h0gauge}a.
Our gauge choice is shown in Fig.~\ref{h0gauge}b, and
details as follows:
\begin{eqnarray*}
t_{R, R+\hat{x}} &=& \sqrt{\frac{3}{8}} \Big(1+ie^{i\pi(x+y)}\Big)
+ \sqrt{\frac{1}{8}} \Big(1-ie^{i\pi(x+y)}\Big) e^{i\pi z} ~, \\
t_{R, R+\hat{y}} &=& \sqrt{\frac{3}{8}} \Big(1-ie^{i\pi(x+y)}\Big)
+ \sqrt{\frac{1}{8}} \Big(1+ie^{i\pi(x+y)}\Big) e^{i\pi z} ~, \\
t_{R, R+\hat{z}} &=& 1 ~.
\end{eqnarray*}
Diagonalizing the kinetic energy, we find two low-energy modes
with normalized real-space wave functions
\begin{eqnarray}
\Psi_1(R) &=& \frac{1 + (\sqrt{3}-\sqrt{2}) e^{i\pi z}}
                         {\sqrt{2(3-\sqrt{6})}}  ~, \\
\Psi_2(R) &=& \frac{1 - (\sqrt{3}-\sqrt{2}) e^{i\pi z}}
                         {\sqrt{2(3-\sqrt{6})}} \times
\frac{e^{i\pi x} - i e^{i\pi y}}{\sqrt{2}} ~.
\end{eqnarray}
Thus, we find two monopole excitations carrying different
lattice momenta.

At the kinetic energy level, any linear combination
\begin{equation}
\Phi(R) = \phi_1 \Psi_1(R) + \phi_2 \Psi_2(R)
\end{equation}
is at the bottom of the monopole band, and there is a continuum 
of states for the monopoles to condense to.
Non-linear terms will lift the degeneracy.  This can be
analyzed near monopole condensation transition by treating 
$\phi_1$ and $\phi_2$ as slowly varying fields and deriving 
Ginzburg-Landau theory of these.

By examining the action of the lattice symmetries,
the resulting Ginzburg-Landau functional is required to be 
invariant under the following transformations
(in our specific gauge):
\begin{eqnarray*}
T_x: & \phi_1 \to \phi_1^* ~, \quad & \phi_2 \to -\phi_2^* ~; \\
T_y: & \phi_1 \to \phi_1^* ~,  \quad & \phi_2 \to \phi_2^* ~; \\
T_z: & \phi_1 \to \phi_2^* ~,  \quad & \phi_2 \to \phi_1^* ~; \\
R_{90^\circ, Rxy}: & \phi_1 \to e^{-i\pi/4} \phi_1^* ~, \quad &
                    \phi_2 \to e^{i\pi/4} \phi_2^* ~; \\
R_{90^\circ, Rxz}: 
& {\displaystyle \phi_1 \to \frac{\phi_1^* + \phi_2^*}{\sqrt{2}} } ~, 
\quad &
  {\displaystyle \phi_2 \to \frac{\phi_1^* - \phi_2^*}{\sqrt{2}} } ~.
\end{eqnarray*}
The $90^\circ$ rotations are about the lattice points on which
monopoles reside.  The hopping Hamiltonian is invariant 
(modulo gauge transformations) under the above lattice symmetries if we 
also complex-conjugate $t_{RR'} \to t^*_{RR'}$ because of the 
``staggered flux'' pattern --- this is the reason for having 
$\phi^*$s on the right hand sides.
There is no direct significance to the seeming ``time-reversal breaking''
in the specific gauge; the physical states of the spin system that we 
study here (with gapped spinons) are time-reversal invariant.

The simplest invariants are $|\phi_1|^2 + |\phi_2|^2$ and
\begin{eqnarray*}
4\left( |\phi_1|^6 |\phi_2|^2 + |\phi_1|^2 |\phi_2|^6 \right) 
- 6|\phi_1|^4 |\phi_2|^4 - (\phi_1^* \phi_2)^4 - (\phi_1 \phi_2^*)^4 ~.
\end{eqnarray*}
The latter can be given a more clear form
\begin{eqnarray}
I_8(\phi_1, \phi_2) &=& N_x^2 N_y^2 + N_y^2 N_z^2 + N_z^2 N_x^2 ~,
\end{eqnarray}
with
\begin{equation}
N_\alpha(\phi_1,\phi_2) \equiv  
{\bm \phi}^\dagger \hat\sigma^\alpha {\bm \phi} ~.
\end{equation}

We write down continuum action for the two-component complex field
${\bm \phi}(R,\tau)$ that respects the above symmetries
\begin{equation}
S_{\rm slow} = \int d\tau d^3 R
\Big[ |(\nabla_\rho - iL_\rho)  {\bm \phi}|^2 
+ U(|{\bm \phi}|^2) + v_8 I_8({\bm \phi}) \Big] + S_L ~,
\label{S2monopole}
\end{equation}
with some potential $ U(|{\bm \phi}|^2) = r|{\bm \phi}|^2 
+ u_4 |{\bm \phi}|^4 + \cdots$.  We have also restored the dual 
gauge field $L_\rho$ and included some generic kinetic energy $S_L$
for $L_\rho$.

We focus on confining paramagnets that obtain starting from
the Coulomb phase and condensing single monopoles so that 
$\la {\bm \phi}\ra \neq 0$; this happens when $r<0$ in the potential.
[It is also possible to have paired condensates such that
$\la {\bm \phi}^\dagger \vec{\sigma} {\bm \phi} \ra \neq 0$ 
but $\la {\bm \phi}\ra = 0$ in some parameter regime --- such states 
correspond to $U(1)$ spin liquids with broken translational symmetry;
these are not considered in any detail here.]
Ground states are selected by minimizing $v_8 I_8({\bm \phi})$,
and the sign of $v_8$ determines the character of the 
resulting phase.

Each state can be characterized by the expectation values 
$N_x, N_y, N_z$ defined above.  
For example, the spatial monopole density in a given state
is given by
\begin{eqnarray}
|\Phi(R)|^2 = |{\bm \phi}|^2 + \frac{1}{\sqrt{3}} 
\Big[ (-1)^x N_x + (-1)^y N_y  + (-1)^z N_z \Big] ~;
\end{eqnarray}
the monopole kinetic energy is
\begin{equation}
E^{(m)}_x = \sqrt{\frac{8}{3}} |{\bm \phi}|^2 
+ \sqrt{2} \Big[(-1)^y N_y  + (-1)^z N_z \Big]
\end{equation}
for $E^{(m)}_x \equiv t_{R,R+\hat{x}} \Phi_R^* \Phi_{R+\hat{x}} + c.c.$,
etc.; while the monopole current is
\begin{equation}
J^{(m)}_x = \sqrt{\frac{2}{3}} 
\Big[(-1)^{x+z} N_y  - (-1)^{x+y} N_z \Big] 
\end{equation}
for $J^{(m)}_x \equiv i(t_{R,R+\hat{x}} \Phi_R^* \Phi_{R+\hat{x}} - c.c.)$.

We also quote transformation properties of 
$\vec{N}=(N_x, N_y, N_z)$ under the lattice symmetries:
\begin{eqnarray*}
& T_x: & (N_x, N_y, N_z) \to (-N_x, N_y, N_z) ~; \\
& T_y: & (N_x, N_y, N_z) \to (N_x, -N_y, N_z) ~; \\
& T_z: & (N_x, N_y, N_z) \to (N_x, N_y, -N_z) ~; \\
& R_{90^\circ, Rxy}: & (N_x, N_y, N_z) \to (N_y, N_x, N_z) ~; \\
& R_{90^\circ, Rxz}: & (N_x, N_y, N_z) \to (N_z, N_y, N_x) ~.
\end{eqnarray*}
We remind that the rotations are about points $R$ of the lattice
on which monopoles reside, which is dual to the original spin model 
lattice.
Given the above information, we can characterize broken lattice 
symmetries when ${\vec N}$ obtains an expectation value.

\subsubsection{$v_8 > 0$}
\begin{figure}
\centerline{\includegraphics[width=3.0in]{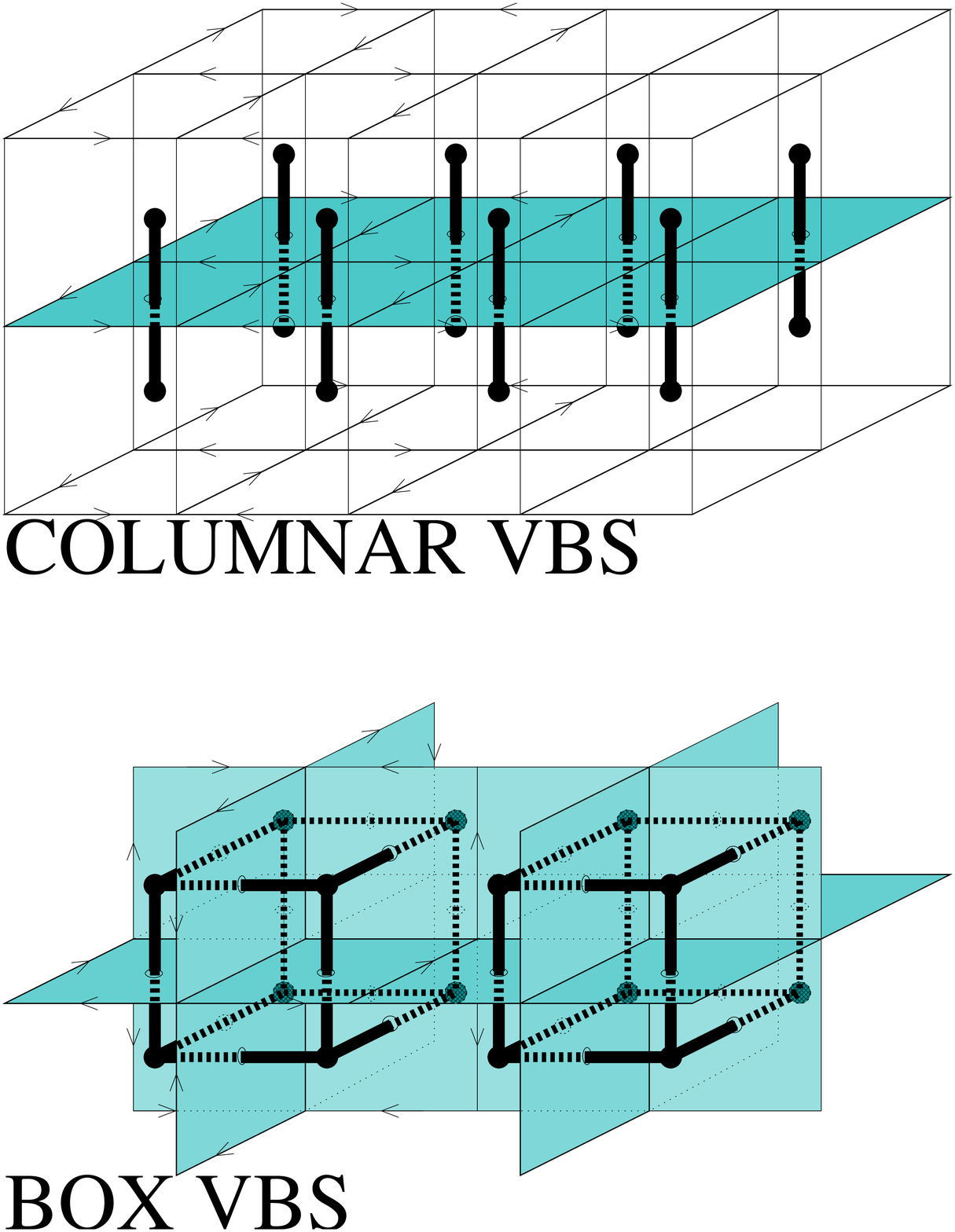}}
\vskip -2mm
\caption{
a) Schematic picture of columnar VBS state obtained
when $v_8>0$.  We have an increased monopole density 
(and also energy density) on the shaded $z$-even planes,
which is interpreted as having valence bonds of the original
spin model preferentially crossing these planes.
Arrows along the placket edges indicate monopole 
currents.
b) 3D box VBS state is obtained when $v_8<0$.
For clarity, only the planes with increased monopole density are shown
($x,y,z$ all even), and the dimers resonate around the cubes
centered where these planes meet.
}
\label{colbox}
\end{figure}

In this case, there are six ground states 
\begin{equation}
(N_x, N_y, N_z)=(\pm 1, 0, 0), \quad (0, \pm 1, 0), \quad (0, 0, \pm 1).
\end{equation}
In a given state, the monopole density is the same in every other 
plane perpendicular to a fixed lattice axis.
For example, the state $(N_x, N_y, N_z)=(0,0,1)$ has an increased 
density on the $z$-even planes and decreased density on the 
$z$-odd planes, and is illustrated in Fig.~\ref{colbox}a.
In the original spin model, these dual planes are crossed by direct
lattice bonds, and on a crude level there is an increased bond
energy crossing the $z$-even planes.
Thus, this state corresponds to a columnar valence bond solid
with dimers oriented in the ${\hat z}$ direction and in the
particular registry.
This identification is also supported by an analysis of the
dual gauge field fluxes induced by the monopole currents
and the connection between these fluxes and the bond 
order parameter.\cite{RSvbs}
The above six states correspond to six possible columnar states
on the cubic lattice.
We expect that this is the phase realized when monopoles condense
in the specific lattice model Eq.~(\ref{Smnpl}), based on the 
analysis of the lattice XY model and the large $\lambda_m$ limit of the
action Eq.~(\ref{Smnpl}).

\subsubsection{$v_8 < 0$}
In this case, there are eight ground states 
\begin{equation}
(N_x, N_y, N_z) = \frac{1}{\sqrt{3}} (\pm 1, \pm 1, \pm 1) ~
\end{equation}
(each of the three signs can be chosen independently).
These states have monopole density oscillating in all three directions
and correspond to three-dimensional ``box'' valence bond solids 
analogous to box states in two dimensions.
For example, the state $(N_x, N_y, N_z) = \frac{1}{\sqrt{3}} (1, 1, 1)$
has maximal monopole density for $x$, $y$ and $z$ all even,
and is illustrated in Fig.~\ref{colbox}b.
In the original spin model, this state has dimers resonating
around direct lattice cubes surrounding these dual lattice points.
The above eight states correspond to eight possible ways to
register such 3D box states.

\subsection{Discussion}
The above considerations lead us to the following picture.
In the spin-1/2 system we expect VBS order parameter to be a 
three-vector
\begin{equation}
\vec{\Psi}_{VBS} =
\left(\begin{matrix} (-1)^x \vec{S}_r \cdot \vec{S}_{r+\hat x} \\
                     (-1)^y \vec{S}_r \cdot \vec{S}_{r+\hat y} \\
                     (-1)^z \vec{S}_r \cdot \vec{S}_{r+\hat z}
\end{matrix}\right) ~.
\end{equation}
We can write down a Ginzburg-Landau functional for $\vec{\Psi}$
\begin{equation}
\label{SVBS}
S_{VBS} = \int d\tau d^3r \Big[ (\nabla_\mu \vec{\Psi})^2 + 
r \vec{\Psi}^2 + u (\vec{\Psi}^2)^2 
+ v (\Psi_x^4 + \Psi_y^4 + \Psi_z^4) \Big]
\end{equation}
where besides $O(3)$-invariant terms we have also included 
quartic interaction with cubic anisotropy $v$, since our
system resides on the cubic lattice.
Thus, the VBS phases in three dimensions are described by an effective 
$O(3)$ model with cubic anisotropy.  Depending on the sign of the 
anisotropy, the three-vector $\vec\Psi$ can either point along 
one of the axes [e.g. $\vec\Psi \propto (0,0,1)$],
which leads to columnar order with 6 degenerate states,
or along a diagonal [e.g. $\vec\Psi \propto (1,1,1)/\sqrt{3}$],
which leads to box order with 8 degenerate states.

The connection with the preceding analysis is provided by
the identification
\begin{equation}
\vec{\Psi}_{VBS} \sim \vec{N} = 
{\bm \phi}^\dagger \vec{\sigma} {\bm \phi} ~.
\end{equation}
Indeed, $\vec\Psi$ has the same transformation properties
under lattice symmetries as $\vec{N}$.
Thus, the three-vector $\vec{\Psi}$ is written with the help of
the two-component complex field ${\bm \phi}$ precisely as in the
$CP^1$ representation of the $O(3)$ nonlinear sigma model. 
As we know, a proper description of the $O(3)$ model in the $CP^1$
language contains besides the spinor field ${\bm \phi}$
also a compact $U(1)$ gauge field, and the monopoles of this
gauge field are conventionally identified with the hedgehog 
configurations of the $O(3)$ vector.  This $CP^1$ gauge field is 
precisely our gauge field $L$ in the ${\bm \phi}$-field action 
Eq.~(\ref{S2monopole}), and as discussed earlier, for a faithful 
description of the original spin-1/2 system we need to allow spin-1/2 
carrying magnetic monopoles in the gauge field $L$,
which correspond to the gapped spinons.

The identification is complete when we realize, extending the discussion 
in  a recent preprint by M.~Levin and T.~Senthil,\cite{avbs} that the 
VBS order parameter field $\vec\Psi_{VBS}$ indeed has hedgehogs 
which are precisely the spinons!  
This is dictated by the microscopic lattice origin of 
$\vec\Psi_{VBS}$, which needs to be brought back into consideration
(this microscopic detail is absent in the continuum action $S_{VBS}$,
Eq.~\ref{SVBS}).
Consider for example the columnar phase and construct a 
hedgehog in the field ${\vec N}$ by putting together the corresponding 
domains of columnar dimer order.  More precisely, we are constructing 
a ``hedgehog'' of the corresponding six-state discrete model.  
When we join these domains, there is an unpaired site 
left which we identify with a spinon residing on one sublattice; 
and if we construct an antihedgehog, we find a spinon residing on the 
other sublattice.  Of course, we pay huge domain wall energies 
consistent with the fact that the spinons are confined in the VBS phase.

\subsection{Connection with the vortex description}
We conclude by describing how to obtain the above picture
directly from the two-vortex action 
Eqs.~(\ref{S2vort})~and~(\ref{includeX0}) at half-filling.
We use hard-spin fields and notation as in 
Appendices~\ref{app:2ch},\ref{app:2ch_halfint}.
Consider phases with gapped chargons obtained by proliferating both 
vortex fields $c^{(1)}$ and $c^{(2)}$.  Ignoring the chargons 
completely, we can expand the cosines containing $\bmnabla \times \bmc$, 
and after simple analysis arrive at the monopole action 
Eq.~(\ref{Smnpl}) with $L_\rho = c^{(1)}_\rho + c^{(2)}_\rho$.
The real utility of the two-vortex action is that it
allows us to incorporate the chargons on equal footing 
in the dual description as ``monopoles'' in the vortex fields 
$c^{(1)}$ and $c^{(2)}$.  
In particular, it formalizes for the easy-plane case the purported 
picture that we are to include monopoles in the gauge field $L_\rho$ 
to obtain the full physical description; no such formalism
is known for the Heisenberg case.

\section{Conclusions}
In this paper, we described a physical mechanism for the appearance
of the fractionalized Coulomb phase in bosonic models.  This was
accomplished by addressing the question of how to view this phase as a 
vortex condensate emanating out of a proximate superfluid phase.
We were led to suggest one route whereby there are two degenerate
but distinct physical vortices in the superfluid phase, and the
superfluid is destroyed by proliferating both species of vortices
with equal amplitude.  The two vortex species correspond to vortices 
whose cores (which are one-dimensional systems) are ordered and 
break a discrete symmetry.  A simple example is provided by bosons at 
half-filling where the vortex core develops insulating checkerboard 
charge density order.  The two vortex species correspond to the two 
checkerboard states in the core. 
Domain walls in the core order survive as gapped excitations when the 
two vortex species proliferate equally.  We argued that these domain
walls may be identified with the monopole excitations of the Coulomb 
phase.  Roton loops formed by combining a vortex loop with one core 
order with an antivortex loop with the other core order were shown to 
correspond to loops of the emergent magnetic flux.  The fluctuations of 
these loops correspond to the gapless photon excitation of the Coulomb 
phase. 

In general, the microscopics that produces degenerate but distinct
physical vortices is model-specific.  But quite broadly, we can view 
these vortices as having ordered cores, where the order is that of some 
nearby insulating phase with a broken discrete symmetry.
For generic bosons at half-filling, one simple candidate is the 
checkerboard charge order, while for the specific corner-sharing 
octahedra model discussed in Sec.~\ref{sec:real} the candidate has 
checkerboard `octahedron' order.

The primary analytical tool used in the paper to put meat into this 
description is a duality mapping to vortex variables which is an 
extension of the familiar boson-vortex duality to (3+1)D.
We extensively developed such a description for the quantum systems of 
interest and used it to discuss the physics.  To discuss the Coulomb 
phases, we showed how the presence of two vortex species can be readily 
incorporated into the dual theory.  The excitation spectrum of the 
Coulomb phase was then derived in this dual formulation which provides 
detailed confirmation of the physical pictures. 
This theory also allows us to identify and describe all basic nearby 
insulating states in a unified setting.

For bosons at half-filling (closely related to spin-$1/2$ quantum 
antiferromagnets), apart from the fractionalized phases, various 
confining phases are possible which break lattice translation symmetry. 
The latter were analyzed by extension to (3+1)D of the methods of 
Haldane,\cite{Haldane} Read, and Sachdev\cite{RSvbs}
(originally developed for lower dimension). 
This  allowed us to discuss possible valence bond orders in this system,
and their relation to the fractionalized Coulomb insulating state.

In the present study of bosonic superfluids and proximate insulators at 
half-filling, we concentrated on vortices with charge-ordered cores.  
It is conceivable instead to have valence bond order inside the core in 
some regime of parameters in the superfluid phase, 
and it would be interesting to develop a broader picture of the 
core dynamics and its consequences.
The above considerations provide further explicit examples of nontrivial
effects arising from the vortex core physics for disordering transitions 
of $XY$ ordered systems, echoing similar phenomena in two-dimensional 
systems.\cite{deccp, avbs} 
In particular, we showed that there can be transformations occurring 
inside the core, and that proliferating ordered cores
can produce unusual phases.  The importance of the core physics
may have broader implications for experimental studies of 
strongly correlated systems.

\acknowledgments
Stimulating discussions with Leon Balents, Mike Hermele, Matthew Fisher,
and Ashvin Vishwanath are gratefully acknowledged.
This research is supported by the NSF grant PHY-9907949
(O.I.M.) and NSF grant DMR-0308945 (T.S.).  T.S also acknowledges 
funding from the NEC Corporation, the Alfred P. Sloan Foundation,
and an award from The Research Corporation.

\appendix

\section{Duality transformation for two-chargon gauge theory 
Eq.~(\ref{S2ch})}
\label{app:2ch}
Here we perform duality analysis of the two-chargon gauge
theory Eq.~(\ref{S2ch}).  Our final result is the two-vortex
action Eq.~(\ref{S2vort}) at integer filling.
We first identify topological defects in the model.
Villain-ized partition function for the model is written as
\begin{widetext}
\begin{eqnarray}
Z_V &=& \int_{-\pi}^{\pi} [D\phi_{1i} D\phi_{2i} Da_{i\mu}] 
\sum_{[p_{1i\mu},\, p_{2i\mu},\, u_{i\mu\nu}] = -\infty}^\infty 
\int_{-\infty}^\infty [Dj_{1i\mu} Dj_{2i\mu} Df_{i\mu\nu}] \; 
e^{-\frac{1}{2\beta} \sum ({\bm j}_1^2 + {\bm j}_2^2)
   -\frac{1}{2K} \sum_{i, \mu<\nu} f_{i\mu\nu}^2} \nonumber\\
& \times &
\exp\{
i\sum \left[{\bm j}_1 \cdot ({\bm \nabla}\phi_1 -{\bm a} - 2\pi{\bm p}_1)
          + {\bm j}_2 \cdot ({\bm \nabla}\phi_2 -{\bm a} - 2\pi{\bm p}_2)
         \right]
  + i\sum_{i, \mu<\nu} f_{i\mu\nu} 
    (\nabla_\mu a_\nu - \nabla_\nu a_\mu - 2\pi u_{i\mu\nu})
\}~.
\label{2chZ}
\end{eqnarray}
\end{widetext}
Real fields ${\bm j}_1, {\bm j}_2, f_{\mu\nu}$ can be interpreted as
the matter currents and the electromagnetic field tensor.

We divide configurations $\{{\bm p}_1, {\bm p}_2, u_{\mu\nu}\}$
into classes equivalent under integer-valued transformations
\begin{eqnarray}
{\bm p}_b^\prime &=& {\bm p}_b + {\bm \nabla} N_b - {\bm V}, 
\quad b=1,2 ~, \nonumber \\
u_{\mu\nu}^\prime &=& u_{\mu\nu} + \nabla_\mu V_\nu - \nabla_\nu V_\mu,
\label{ivalgauge}
\end{eqnarray}
with integer fields $N_1, N_2, {\bm V}$.
These classes are specified by two vorticities
\begin{equation}
q_{\mu\nu}^{(b)} = \nabla_\mu p_{b\,\nu} - \nabla_\nu p_{b\,\mu} 
+ u_{\mu\nu} ~,
\end{equation}
and monopole currents
\begin{equation}
J_{\rho}^{(m)} = \frac{1}{2} \epsilon_{\rho\sigma\mu\nu} 
\nabla_\sigma u_{\mu\nu} ~.
\end{equation}
The latter satisfy continuity
\begin{equation}
\nabla_\rho J_\rho^{(m)} = 0 ~,
\label{Jmconstr}
\end{equation}
while the vorticities satisfy $dq^{(b)} = du$, which can be written as an 
integer-valued constraint
\begin{equation}
\frac{1}{2} \epsilon_{\rho\sigma\mu\nu} 
\nabla_\sigma q_{\mu\nu}^{(b)} = J_{\rho}^{(m)}, \quad b=1,2.
\end{equation}
The meaning of the last equation is that monopoles act as sources and 
sinks for both vorticities simultaneously.

For each allowed configuration of vorticities and monopole currents,
we can now perform the summation over all 
$\{{\bm p}_1, {\bm p}_2, u_{\mu\nu}\}$ in the corresponding
class using Eqs.~(\ref{ivalgauge}).  This effectively extends
the integration variables $\phi_1, \phi_2, a_\mu$ over the 
whole real line, and the integrals give the delta function 
conditions
\begin{eqnarray*}
[\delta(\bm{\nabla} \cdot \bm{j}_1)] [\delta(\bm{\nabla} \cdot \bm{j}_2)]
[\delta(\nabla_\nu f_{\mu\nu} - j_{1\mu} - j_{2\mu})] \propto \\
\propto [\delta(\bm{\nabla} \cdot \bm{j}^-)] 
        [\delta(j^{+}_\mu - \nabla_\nu f_{\mu\nu})] ~.
\end{eqnarray*}
In the last line, we changed variables
to ${\bm j}^{\pm} = {\bm j}_1 \pm {\bm j}_2$.
Note that not all conditions in the first line are independent,
and the precise meaning of the delta functions is given by the 
second line.  Thus, we can completely eliminate ${\bm j}^{+}$.

The topological defects enter through
\begin{eqnarray*}
i \pi \sum ({\bm j}^{+} ({\bm p}_1 + {\bm p}_2)
            + {\bm j}^{-}  ({\bm p}_1 - {\bm p}_2))
+ i 2\pi \sum_{\mu<\nu} f_{\mu\nu} u_{\mu\nu} \\
= i \pi \sum_{\mu<\nu} f_{\mu\nu} (q_{\mu\nu}^{(1)} + q_{\mu\nu}^{(2)})
+ i \pi \sum {\bm j}^{-}  ({\bm p}_1 - {\bm p}_2) ~.
\end{eqnarray*}
We now solve the constraint ${\bm \nabla} \cdot {\bm j}^{-}=0$ 
with a rank-2 field $\g_{I\rho\sigma}^{-}$ defined on the 
dual plackets
\begin{equation}
j^{-}_\mu = \frac{1}{2} \epsilon_{\mu\nu\rho\sigma} 
\nabla_\nu \frac{\g_{\rho\sigma}^{-}}{\pi} ~.
\label{gm2jm}
\end{equation}
It is also convenient to pass from the variables $f_{\mu\nu}$ to their 
dual $\g_{\rho\sigma}^{+}$ via
\begin{equation}
f_{\mu\nu} = \frac{1}{2} \epsilon_{\mu\nu\rho\sigma} 
\frac{\g_{\rho\sigma}^{+}}{\pi} ~;
\end{equation}
${\bm j}^+$ is obtained from $\g^+$ by an expression similar to 
Eq.~(\ref{gm2jm}).
Finally, it is convenient to specify vorticities
$q^{(b)}_{\mu\nu}$ by the corresponding integer-valued fields 
$F^{(b)}_{\rho\sigma}$, $b=1,2$,
\begin{equation}
F^{(b)}_{\rho\sigma} = \frac{1}{2} \epsilon_{\rho\sigma\mu\nu} 
q^{(b)}_{\mu\nu} ~, \quad
\nabla_\sigma F^{(b)}_{\rho\sigma} = J^{(m)}_\rho ~,
\label{Fconstr}
\end{equation}
just as we did for the (3+1)D XY model in Eq.~(\ref{q2F}).

Putting everything together, the partition sum reads
\begin{widetext}
\begin{eqnarray}
Z = \sum_{[F^{(1)}_{I\rho\sigma},
           F^{(2)}_{I\rho\sigma}, J^{(m)}_{I\rho}]}^\prime
\int_{-\infty}^\infty [D\g_{I\rho\sigma}^{+}] 
                      [D\g_{I\rho\sigma}^{-}]^\prime \;
e^{-\frac{1}{4\beta} \sum ({\bm j}_+^2 + {\bm j}_-^2)
   -\frac{1}{2K \pi^2} \sum_{\rho<\sigma} (\g_{\rho\sigma}^{+})^2}
e^{- i\sum_{\rho<\sigma} 
      [\g_{\rho\sigma}^{+} (F^{(1)} + F^{(2)})_{\rho\sigma}
       + \g_{\rho\sigma}^{-} (F^{(1)} - F^{(2)})_{\rho\sigma}] } ~.
\label{2chZtop}
\end{eqnarray}
\end{widetext}
We can now in principle integrate out the fields $\g^{+}_{\rho\sigma}$
and $\g^{-}_{\rho\sigma}$, and obtain a theory in terms of vortex 
worldsheets with sources and sinks on the monopole worldlines.
The vortex interactions are such that combinations 
$F^{(1)} + F^{(2)}$ enter with short-ranged interactions similar
to screened vortices, while $F^{(1)}-F^{(2)}$ enter with 
long-ranged interactions of unscreened vortices.  This is expected
since the chargon field combination $\phi_1 + \phi_2$ is gauged
while $\phi_1-\phi_2$ is gauge neutral.
We expect the physics to remain unchanged upon adding local
vortex and monopole fugacity terms
\begin{equation}
S_{\rm fug.} = \frac{1}{2\lambda} \sum_{I,\rho<\sigma}
[ (F^{(1)}_{I\rho\sigma})^2 + (F^{(2)}_{I\rho\sigma})^2 ]
+ \frac{1}{2\lambda_m} \sum_{I\rho} (J^{(m)}_{I\rho})^2 ~.
\end{equation}
Just as in the case with the $XY$ model in Sec.~\ref{subsec:xy}, 
to get a better intuition about the theory at hand, we instead 
introduce two compact $U(1)$ gauge fields 
$c_{I\rho}^{(1)}$, $c_{I\rho}^{(2)}$, and a $U(1)$ scalar field 
$\theta^{(m)}_I$ that implement the constraints 
Eqs.~(\ref{Fconstr},\ref{Jmconstr}) and are the appropriate 
conjugate variables.

The final expression reads
\begin{widetext}
\begin{eqnarray}
Z_V[\lambda, \lambda_m] = 
\int_{-\pi}^\pi [Dc^{(1)}_{I\rho} Dc^{(2)}_{I\rho} D\theta^{(m)}_I]
\int_{-\infty}^\infty [D\g_{I\rho\sigma}^{+}]
                      [D\g_{I\rho\sigma}^{-}]^\prime 
e^{-\frac{1}{4\beta} \sum ({\bm j}_+^2 + {\bm j}_-^2)
   -\frac{1}{2K \pi^2} \sum_{\rho<\sigma} (\g_{\rho\sigma}^{+})^2}
\nonumber \\
\times  \exp[
\lambda_m \sum_{I\rho} \cos(\nabla_\rho \theta^{(m)} - c_{I\rho}^{(1)} 
                                                     - c_{I\rho}^{(2)})
+ \lambda \sum_{I, \rho<\sigma} 
\cos(\nabla_\rho c_\sigma^{(1)} - \nabla_\sigma c_\rho^{(1)}
     - \g_{I\rho\sigma}^{+} - \g_{I\rho\sigma}^{-} )
\nonumber \\
+ \lambda \sum_{I, \rho<\sigma} 
\cos(\nabla_\rho c_\sigma^{(2)} - \nabla_\sigma c_\rho^{(2)}
     - \g_{I\rho\sigma}^{+} + \g_{I\rho\sigma}^{-} )
] ~.
\label{dualS2ch}
\end{eqnarray}
\end{widetext}
As usual, cosines stand for the appropriate Villain forms.
At this stage, the field $\g^+_{\rho\sigma}$ is massive and can be 
integrated out.  The variables of the resulting dual theory are:
i) two compact $U(1)$ gauge fields $e^{i c^{(1)}}$ and $e^{i c^{(2)}}$, 
which can be viewed as vortex line segment creation operators;
ii) monopole matter field $e^{i \theta^{(m)}}$ that carries both 
dual gauge charges;
iii) noncompact rank-2 field $\g^-$ that further ``gauges'' 
the vortex fields $c^{(1)}$ and $c^{(2)}$ and describes the
physical boson density fluctuations.
Equation~(\ref{S2vort}) in Sec.~\ref{subsec:2vort} displays a 
`soft-spin' version of this dual theory by writing
\begin{equation}
\Psi^{(1)*}_{I\rho} \sim e^{ic^{(1)}_{I\rho}}, \quad 
\Psi^{(2)*}_{I\rho} \sim e^{-ic^{(2)}_{I\rho}}, \quad
\Upsilon_I^* \sim e^{i\theta^{(m)}_I}
\end{equation}
(we also dropped the superscript on the rank-2 gauge field $\g^-$ 
that survives in the low energy theory). 

The analysis of the possible phases from this dual perspective 
can be done as in the XY case in Sec.~\ref{subsec:xy} 
by first considering a theory without the $\g^-$ field
(i.e., ``screened vortex'' theory which would obtain if the two 
chargons were also coupled to a fluctuating external 
electromagnetic field.)
The field $\g^-$ is then included semiclassically and
provides correct count of the low-energy modes.

Thus, the screened vortex theory has a phase for 
$\lambda \gg 1$, $\lambda_m \ll 1$ in which both fields $c^{(1)}$ and 
$c^{(2)}$ are deconfined while the monopole field $\theta^{(m)}$ is 
gapped.  The screened vortex theory has two photons in this phase.
Including fluctuations in the rank-2 field $\g^-$, it obtains a mass 
and eats in the process one photon, so there remains precisely one 
photon $[\nabla \times (c^{(1)}+c^{(2)})]^2$.
In the physical boson model, this is the fractionalized Coulomb 
phase at the focus of this paper.

On the other hand, in the regime $\lambda \gg 1$, $\lambda_m \gg 1$
monopoles are also proliferated and kill the photon, so this 
becomes the conventional Mott insulator of bosons.

Finally, for small $\lambda \ll 1$ both dual gauge fields $c^{(1)}$ and 
$c^{(2)}$ are confined.  The dual electric field lines which are the 
vortices of the physical boson model are gapped, and this corresponds
to the superfluid phase of bosons.
As we vary the parameter $\lambda_m$ inside this phase, there is a 
transition in the properties of these line excitations from a unique 
core state to two degenerate core states.  

The dual formulation reveals some details of this core transition.
Specifically, we consider a single vortex in the screened vortex 
theory --- this allows us to avoid complications of coupling the vortex 
dynamics to phonons and focus on the core physics.

It is convenient to use Hamiltonian formulation
\begin{eqnarray*}
\label{}
H_{\rm screened} = u_m \sum n_R^2 
- t_m \sum \cos(\nabla\theta^{(m)} - \bmc^{(1)} - \bmc^{(2)}) \\
+ u_v \sum (\bme_1^2  +  \bme_2^2) 
- t_v \sum \left[\cos(\nabla \times \bmc^{(1)}) 
                 + \cos(\nabla \times \bmc^{(2)}) \right] ~,
\end{eqnarray*}
with the Hilbert space constraints
\begin{equation}
\bmnabla \cdot \bme_1 = \bmnabla \cdot \bme_2 = n_R ~.
\end{equation}
Here $u_m$ and $t_m$ are bare monopole gap and hopping amplitude,
while $u_v$ is bare vortex core energy per unit length and
$t_v$ is vortex hopping amplitude.
The superfluid phase of the boson model corresponds to the
the confining phase $u_v \gg t_v$ of $H_{\rm screened}$.
The electric field $\bme \equiv \bme_1 - \bme_2$ is the conserved 
physical vorticity, and we study a strength-1 line of $\bme$ 
oriented in say $\hat x$ direction.
In the limit $u_v \gg t_v$, we have a straight line $e_x=1$, 
but its segments can be realized as either 
$(e_{1x},e_{2x})=(1,0)$ or $(0,-1)$.
Each juncture $(0,-1) \to (1,0)$ has a $+1$ monopole and
$(1,0) \to (0,-1)$ has a $-1$ monopole, and the quantum 
dynamics is determined by the competition between the $u_m$ and $t_m$
terms.  Writing $\sigma_l^z = +1$ for $(1,0)$ segment and
$\sigma_l^z = -1$ for $(0,-1)$ segment, we obtain the
following 1D Hamiltonian for the straight line
\begin{equation}
H_{\rm core} = -\frac{u_m}{2} \sum_l \hat\sigma_l^z \hat\sigma_{l+1}^z
- \frac{t_m}{2} \sum_l \hat\sigma_l^x ~.
\end{equation}
This is simply the one-dimensional quantum Ising model,
and has two phases:
For $u_m < t_m$, we obtain a paramagnet in $\sigma^z$ and 
the ground state is unique, schematically $\sigma^x = 1$.
In this case, the monopoles proliferate along the line,
and the $e=1$ line is unique.
On the other hand, for $u_m > t_m$, we obtain a ferromagnet with
two degenerate ground states $\sigma^z = +1$ and $\sigma^z = -1$.
The monopoles remain gapped along the line, and we have two 
distinct $e=1$ lines.  The core Hamiltonian can be also written in 
dual Ising variables, and this gives a description in terms of 
domain wall particles which we identify with the monopoles.

The above analysis is valid in the limit $u_v \gg t_v$,
but we expect the two regimes and the transition between them
to extend throughout the confined phase of $H_{\rm screened}$.
In the superfluid phase of the original boson model, we conjecture 
that the corresponding vortex core transition line extends all the way 
to the insulating phases as shown in Fig.~\ref{cqedw2m_phased}.

\section{Duality analysis for bosons at half-integer filling}
\label{app:halfint}
Here we generalize the analysis in Sec.~\ref{subsec:xy} to 
bosons at half-integer filling and study the resulting dual theory
directly.  At half-filling, we simply make the replacement 
\begin{equation}
j_{i\tau}^2 \to \left(j_{i\tau}-\frac{1}{2}\right)^2
\end{equation}
in the final expression Eq.~(\ref{xydual:final}) while the rest 
remains unchanged.  
As in (2+1)D, noninteger boson density enters as an external
``field'' seen by the vortices, except that vortices are now
lines and the gauge potential is rank-2 tensor.

A convenient formulation is obtained by finding static 
$\g^0_{\rho\sigma}$ such that
\begin{equation}
\frac{1}{2} \delta_{\mu\tau}
= \frac{1}{2} \epsilon_{\mu\nu\rho\sigma} 
\nabla_\nu \frac{\g^0_{\rho\sigma}}{2\pi} ~.
\end{equation}
If $\g^0_{\rho\sigma}$ are viewed as fluxes through the dual lattice 
plackets, $\g^0$ has total flux of $\pi$ coming out of each 
spatial cube.  Shifting $\g$ by $\g^0$ we get
\begin{eqnarray*}
Z = && \int_{-\infty}^\infty [D{\g_{I\rho\sigma}}]^\prime
\int_{-\pi}^\pi [Dc_{I\rho}]  
\exp\left[-\frac{1}{2\beta} \sum (d\g)^2 \right]
\nonumber
\\
&& \times \exp\left[ \lambda \sum_{I, \rho<\sigma} 
\cos(\nabla_\rho c_\sigma - \nabla_\sigma c_\rho - \g_{I\rho\sigma}
     - \g^0_{I\rho\sigma}) \right]. 
\end{eqnarray*}
Thus, there is a static contribution to the rank-2 gauge potential 
corresponding to placing half of a magnetic monopole inside each 
spatial cube.  When a vortex line evolves in space-time, 
this gives a Berry phase contribution of $\pi$ for each cube in the 
3D volume swept by the line.

As in our discussion for bosons at integer filling, it is
helpful to first consider the compact $U(1)$ part with only 
the static $\g^0_{\rho\sigma}$.
This ``frustrated'' lattice gauge theory has not been studied
to our knowledge.  At present, we do not have direct tools for 
approaching this problem similar to the ones used for 
frustrated systems with global symmetries.
We can still describe what we expect to happen in such a model.

Two phases of the frustrated lattice gauge theory are clearly 
identified.  
For small $\lambda$, we expect confinement in $e^{ic_{I\rho}}$, 
which corresponds (upon restoring fluctuations in $\g_{\rho\sigma}$)
to the superfluid phase of the bosons.

For large $\lambda$, we expect deconfinement in $e^{ic_{I\rho}}$,
but we need to be more specific since there can be different
patterns of deconfinement in the presence of $\g^0_{\rho\sigma}$.
For the `hard-spin' action written above, we can proceed
by considering first the classical ground states and obtain the 
following picture.
For each spatial cube, the total outgoing flux of 
$e^{i(\nabla_\rho c_\sigma - \nabla_\sigma c_\rho - \g^0_{\rho\sigma})}$
must be $\pi$ modulo $2\pi$ because of the compactness of the
gauge field.  For an individual cube, the lowest energy is obtained by 
dividing this flux equally among the six faces.
On the lattice, we put such cubes with the outgoing placket fluxes of 
$+\pi/6$ and cubes with the outgoing fluxes of $-\pi/6$ in a checkerboard
pattern, and there are two degenerate ground states.
This structure survives as a phase in the frustrated lattice gauge 
theory.
For the original bosonic system, this phase corresponds to 
checkerboard charge density wave.  Indeed, when we allow
fluctuations in $\g_{\rho\sigma}$, the above staggered
character is imprinted on the fluxes $\g_{\rho\sigma}$,
thus producing staggered boson density.
The two states correspond to two ways to register this 
CDW on the lattice.

Other deconfinement patterns are also possible, but likely 
require some deformation of the above simple action
to be stabilized.
For example, it is possible to have a deconfined state with an 
energy density wave, and this would correspond to some
VBS phase for the original bosons.

Of direct interest here is the possibility of deconfinement in 
$e^{ic_{I\rho}}$ that produces {\em two} photons in the 
low-energy description and does not break any lattice symmetry.  
Allowing fluctuations in $\g_{\rho\sigma}$, only one photon remains, 
and this would give the sought for fractionalized Coulomb phase at 
half-integer filling.
We know that this can indeed happen in the original bosonic system,
and Appendices~\ref{app:2ch} and \ref{app:2ch_halfint} provide
an indirect route to describe such a phase.
It would be interesting to explore this in the above frustrated 
lattice gauge theory in more detail and develop direct analytical 
tools for treating such theories.

\section{Duality for two-chargon theory at half-integer filling}
\label{app:2ch_halfint}
In this Appendix, we derive dual description for a two-chargon gauge 
theory corresponding to bosons at half-integer filling.
Our analysis is a direct extension to (3+1)D of Appendix E of 
Ref.~\onlinecite{deccp}, which should be consulted for further details.
Following Ref.~\onlinecite{deccp}, we arrive at a path integral 
which has the same form as at integer filling Eq.~(\ref{2chZ}) 
but with replacements
\begin{eqnarray*}
\left(j_{1i\tau}\right)^2 \to \left(j_{1i\tau}-n_0 \eta_i\right)^2, \quad
\left(j_{2i\tau}\right)^2 \to \left(j_{2i\tau}-n_0 \eta_i\right)^2 ~,
\end{eqnarray*}
and an additional term in the action
\begin{equation}
S_B = i \sum_i \eta_i a_{i\tau} ~.
\end{equation}
In the limit $K \to 0$ (i.e. in the absence of the gauge field dynamics),
any choice of the ``chemical potential'' $n_0$ for the on-site gauge
charge will give the same result due to constraint 
$j_{1i\tau} + j_{2i\tau} = \const$.  
This is no longer true for finite $K$, but the qualitative behavior 
is expected to be insensitive to $n_0$, and we will use this freedom 
to make the structure of the dual theory more transparent.

We proceed as in Appendix~\ref{app:2ch} while treating the phase term 
$S_B$ as in our analysis of the monopole Berry phase for spin-1/2
systems in Sec.~\ref{sec:vbs}.  Specifically, we take $f^0_{\mu\nu}$ 
defined by conditions Eq.~(\ref{f0:1},\ref{f0:2})
and make the corresponding change of variables
$f_{\mu\nu} = \tilde{f}_{\mu\nu} + f^0_{\mu\nu}$.
In terms of the variables $j_{1\mu}, j_{2\mu}, \tilde{f}_{\mu\nu}$,
we obtain the same action as at integer filling except for 
additional contributions
\begin{equation}
\frac{1}{K} \sum_{\mu<\nu} \tilde{f}_{\mu\nu} f^0_{\mu\nu}
- \frac{n_0}{\beta} \sum \eta_i j^+_{i\tau} 
+ i\sum_{\mu<\nu} f^0_{\mu\nu} 2\pi u_{\mu\nu} ~.
\end{equation}
In particular, upon integrating out $\phi_1,\phi_2,a_\mu$,
we have $j^+_\mu = \nabla_\nu \tilde{f}_{\mu\nu}$.
The first two terms cancel each other if $Kn_0/\beta= 1/12$
(see Eq.~\ref{Sn0}).
The importance of the specific form for $f^0_{\mu\nu}$
is that for general $n_0$ the first two terms add up to
$\frac{\delta n_0}{\beta} \sum_i \eta_i j^+_{i\tau}$, which can
be viewed as a staggered chemical potential seen by the gauge charges.
As long as this potential is small, it effectively averages out and 
does not modify the low-energy modes; in the following, we set 
$\delta n_0 = 0$.

The remaining phase term can be expressed entirely in terms of the 
monopole currents $J^{(m)}_\rho$
\begin{equation}
\tilde{S}_B = i \sum J^{(m)}_\rho X^0_\rho ~,
\end{equation}
with $X^0_\rho$ defined in Section~\ref{sec:vbs}.
This term needs to be added to the dual action expressed in terms 
of the topological defects Eq.~(\ref{2chZtop}).
Thus, we have managed to absorb all effects of half-integer filling
into the monopole Berry phases.
To obtain the final action in terms of the dual fields
$\theta^{(m)}, c^{(1)}, c^{(2)}$, we simply take the expression
at integer filling Eq.~(\ref{dualS2ch}) and make the replacement
\begin{equation}
\nabla_\rho \theta^{(m)} \to \nabla_\rho \theta^{(m)} - X^0_\rho ~.
\label{includeX0}
\end{equation}

\section{Ising P$^*$ phase in (2+1)D}
\label{app:Ising}
\begin{figure}
\centerline{\includegraphics[width=2.6in]{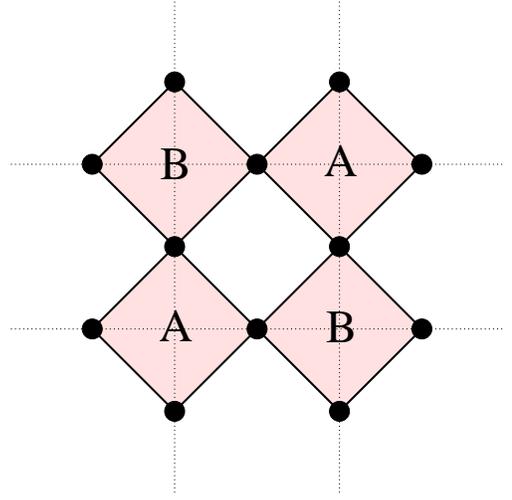}}
\vskip -2mm
\caption{Quantum Ising model that realizes $P^*$ phase in two 
dimensions.  Spins are located on the link-centered sites of a 
square lattice.  Each shaded diamond indicates the $U$-term
in the Hamiltonian Eq.~(\ref{HIsing}).
}
\label{diamonds}
\end{figure}

{\bf Toy model:}
Here we consider a toy model which is a 2D quantum Ising version of the 
3D boson model of Sec.~\ref{sec:real}.  As we will see, some physical
aspects are similar in the two problems, while the formalism is much
simpler in the Ising case and gives some guidance for the bosons.
Ising spins reside on links of a square lattice, and the Hamiltonian is
\begin{equation}
H = -J \sum_{\la ll' \ra} \sigma^z_l \sigma^z_{l'}
- \Gamma \sum_l \sigma^x_l
- U \sum_r \prod_{l \in r} \sigma^x_l ~.
\label{HIsing}
\end{equation}
The lattice is shown in Fig.~\ref{diamonds}.
The spins can be viewed as forming a network of corner-sharing 
diamonds, and we introduced four-spin interaction $U$ on each diamond.
For simplicity, we take $U>0$.

The phase diagram of the model can be analyzed as in 
Sec.~\ref{sec:real}.  When the Ising coupling dominates,
$J \gg \Gamma,U$, the system is ferromagnetically ordered.
When the transverse fields dominate, $\Gamma, U \gg J$,
the system is a conventional Ising paramagnet.

The intermediate regime $U \gg J \gg \sqrt{\Gamma U}$ is most
interesting and realizes an unusual topologically ordered paramagnet 
(Ising P$^*$ phase).
The excitation spectrum in this phase contains a gapped Ising vortex 
(vison) and gapped Ising matter fields.  
The argument is similar for the Ising and boson models.  
In the large $U$ limit, the ground state sector is determined by the 
projection $\prod_{l \in r} \sigma_l^x = 1$ for all $r$, and the 
effective Hamiltonian in this sector is
\begin{eqnarray}
\label{Heff:Ising}
H_{\rm eff}^{(0)} = - \Gamma \sum_l \sigma^x_l
-K \sum_\Box \sigma^z_{12} \sigma^z_{23} \sigma^z_{34} \sigma^z_{41} 
\end{eqnarray}
with $K=J^2/U$.  This is just the familiar Ising gauge theory.
For $K \lesssim \Gamma$ the theory is confining and corresponds to the 
conventional paramagnet, but for $K \gtrsim \Gamma$ we have a 
deconfined phase and this is the Ising P$^*$ phase.

The topology of the phase diagram is the same as in Fig.~\ref{phased},
with the correspondences superfluid $\to$ ferromagnet,
conventional insulator $\to$ conventional paramagnet, and 
fractionalized insulator $\to $ P$^*$ paramagnet.
We are led to ask similar questions on the relationships among the 
phases, in particular, how to view the P$^*$ phase coming from the
ferromagnet.  In the ferromagnet, excitations are domain walls,
and we can think of a paramagnet as a result of proliferating these.
The P$^*$ phase features the vison, which is a topological point 
excitation, and we ask how this appears in the domain wall condensate.

To push the analogy somewhat further, in (3+1)D the appearance 
of the Coulomb phase with the monopole excitation is `natural' when an 
$O(3)$ spin model is disordered without proliferating hedgehogs
(gapped hedgehogs then become the monopoles).
As discussed in the main text, the appearance of the Coulomb phase 
in a nominally $O(2)$ model is more subtle in terms of the 
vortex line defects of the ordered phase.  
Similar situation occurs in (2+1)D, where the appearance of the
$Z_2$ fractionalized phase with the vison excitation is natural
when an $O(2)$ model is disordered by proliferating double
vortices but not single vortices, but the appearance of such
$Z_2$ phase is not so clear in a nominally Ising model.
We argue that the key to this puzzle lies in the presence of
two degenerate but distinct domain walls in the ferromagnetic
phase near the P$^*$ phase.

{\bf Direct analysis of domain walls:}
In the present model, we can perform a direct analysis in the 
limit of small $\Gamma$, i.e., in the upper right corner of the 
phase diagram Fig.~\ref{phased}.

When $\Gamma=0$ in the microscopic model, we find that for each square 
lattice placket the corresponding $Z_2$ `flux'
$\Phi_P \equiv \sigma^z_{12} \sigma^z_{23} \sigma^z_{34} \sigma^z_{41}$ 
is conserved.  This simplifies the analysis, since in each sector 
with fixed $\Phi$'s the Hamiltonian decouples into two quantum Ising 
models.  Indeed, let us work in the $\sigma^z$ basis and fix a
sector $\{\Phi_P\}$.  Let us find a specific realization 
$\{\sigma^{z(0)}_{rr'}\}$ of the fluxes, 
$\Phi_P =  \sigma^{z(0)}_{12} \sigma^{z(0)}_{23} 
           \sigma^{z(0)}_{34} \sigma^{z(0)}_{41}$.
$\{\sigma^z_{rr'}\}$ belongs to this sector if we can write
$\sigma^z_{rr'} = S^z_r S^z_{r'} \sigma^{z(0)}_{rr'}$,
and we can label all states in the sector by the new Ising variables
$\{S^z_r\}$.  The action of the Hamiltonian in this sector reads
\begin{eqnarray*}
H_\Phi &=& -\sum_{\la AA' \ra} J_{AA'} S^z_A S^z_{A'} - U\sum_A S^x_A \\
&&         -\sum_{\la BB' \ra} J_{BB'} S^z_B S^z_{B'} - U\sum_B S^x_B ~.
\end{eqnarray*}
Here $A$ and $B$ refer to the two sublattices of the square lattice.
Ising couplings $J_{AA'}$ and $J_{BB'}$ depend on the sector 
$\{\Phi_P\}$; 
for example, $J_{AA'}$ is determined by considering the square lattice 
placket $[1_A, 2, 3_{A'}, 4]$ that has $AA'$ as its diagonal:
$J_{AA'} = J(\sigma^{z(0)}_{12} \sigma^{z(0)}_{23}
             + \sigma^{z(0)}_{14} \sigma^{z(0)}_{43})$.

For arbitrary $J/U$ (with $\Gamma=0$) the ground state sector has no 
fluxes, $\Phi_P = 1$, and $J_{AA'}=J_{BB'}=2J$.
As we change $J/U$, we have a simultaneous Ising ordering transition for 
spins $S_A$ and $S_B$.  The gap to the nearest sector is at least of
order $J$ in the ferromagnetic phase and of order $J^2/U$ in the 
spin-disordered phase.

Consider the ferromagnetic phase.  In the sector with no fluxes,
excitations are domain walls for the Ising variables $S_A$ or $S_B$.  
When translated to the original variables $\sigma^z_{rr'}$, 
a domain wall in $S_A$ is a physical domain wall that passes through the 
$B$ sublattice sites, while a domain wall in $S_B$ is a physical domain 
wall that passes through the $A$ sites.
In a sector with two visons, one finds that the lowest energy state
has an $A$-type domain wall (defined by frustrated bonds 
$J_{AA'} S^z_A S^z_{A'}$) connecting the two visons, and also a 
$B$-type domain wall connecting the visons.  
Thus, visons act as sources for both domain walls and are linearly 
confined.

Consider now the spin-disordered phase.  It can be viewed as a 
condensate of both domain walls, and has two Ising
matter fields $S_A$ and $S_B$ with the mass of order $U$.
The different sectors are accounted for as vison excitations with
the vison gap of order $J^2/U$.

For nonzero $\Gamma$, the different sectors will mix, but the above 
description of the phases will remain true as long as $\Gamma$ 
is much smaller than the gap to the nearest sector.  
Small nonzero $\Gamma$ will induce energy-energy coupling between the 
two Ising models; this will affect the nature of the transition, either 
making it first-order or driving to a different universality class;
however, the above physical picture with two distinct domain walls 
remains.

{\bf Phenomenological gauge theory description} similar to the 
two-chargon theory is also possible.  Again, it relies on the 
observation that in the P$^*$ phase we find two distinct Ising matter 
excitations.  The phenomenological Hamiltonian is
\begin{eqnarray*}
H_{\rm 2is} &=& -U\sum_r (\tau^x_{1r} + \tau^x_{2r})
- J\sum_{\la rr' \ra} (\sigma^z_{rr'} \tau^z_{1r} \tau^z_{1r'}
                       + \sigma^z_{rr'} \tau^z_{2r} \tau^z_{2r'}) \\
&& -\Gamma \sum_{\la rr' \ra} \sigma^x_{rr'} 
- K\sum_\Box \sigma^z_{12} \sigma^z_{23} \sigma^z_{34} \sigma^z_{41} ~;
\end{eqnarray*}
it contains two Ising matter fields $\tau_{1r}, \tau_{2r}$
coupled to an Ising gauge field $\sigma_{rr'}$.  The Hilbert space 
of the theory is defined by
\begin{equation}
\tau_{1r}^x \tau_{2r}^x \prod_{r'\in r} \sigma_{rr'}^x = 1 ~.
\end{equation}
We emphasize that this model is not derived from the Hamiltonian 
Eq.~(\ref{HIsing}) (in particular, the reader should not be confused
by our reuse of the letters for the coupling constants).  
Rather, this model is to be considered as capturing the 
relevant physics of the P$^*$ phase.

The analysis of the phenomenological gauge theory proceeds exactly
as in the two-chargon theory, and the phase diagram has the same 
topology as in Fig.~\ref{cqedw2m_phased}.
The ordered phase has ferromagnetic order in the `physical' 
spin $\tau^z_r \equiv \tau^z_{1r} \tau^z_{2r}$.  In the
ferromagnetic phase, we have domain wall excitations, and
a domain wall segment crossing a link $\tau^z_r \tau^z_{r'} = -1$ 
can be realized as either a domain wall in $\tau_1$ 
($\sigma^z_{rr'} \tau_{1r}^z \tau_{1r'}^z = -1$) or a domain 
wall in $\tau_2$.  $Z_2$ flux (vison) acts as a source for both
domain walls, and there is a linearly confining potential
between a pair of visons in the ferromagnetic phase.  
On a physical domain wall, however, vison costs only finite energy and 
can hop along the domain wall;
this hopping represents quantum tunneling between the two microscopic 
realizations of the hopped segment.  Vison can be thought of 
as a point `domain wall' for the order along the line which is the 
Ising domain wall of the ferromagnetic phase.

We expect two possibilities:  One is that visons proliferate 
along the domain wall, in which case the domain wall is unique.
The other possibility is that visons remain gapped inside the 
domain wall, and there are two distinct domain walls.
A precise formulation is to consider an externally imposed
domain wall in the ferromagnetic phase.  As we change the
parameters in the ferromagnetic phase, there is a transition from a 
disordered domain wall with a unique ground state to an ordered 
domain wall with two degenerate ground states, and this domain wall 
ordering transition is of (1+1)D Ising type.
When we have ordered domain walls and proliferate both with equal
amplitudes, the P$^*$ phase results.

The mathematical formalism for the above picture is transformation
to dual variables, which is readily done in the Hamiltonian language.
The dual Hamiltonian is
\begin{eqnarray*}
H_{\rm 2is, dual} &=& 
-U\sum_\Box (C^z_{12} C^z_{23} C^z_{34} C^z_{41}
             + D^z_{12} D^z_{23} D^z_{34} D^z_{41}) \\
&& - J\sum_{\la RR' \ra} (C^x_{RR'} + D^x_{RR'}) \\
&& - \Gamma \sum_{\la RR' \ra} V_R^z V_{R'}^z C^z_{RR'} D^z_{RR'}
- K\sum_R V^x_R ~;
\end{eqnarray*}
the Hilbert space is defined by
\begin{equation}
V^x_R \prod_{R'\in R} C^x_{RR'}=1, \quad
V^x_R \prod_{R'\in R} D^x_{RR'}=1.
\end{equation}
The two domain wall variables are denoted as $C, D$, 
while the $Z_2$ flux variable is denoted as $V$:
$C^x_{RR'} \equiv \sigma^z_{rr'} \tau^z_{1r} \tau^z_{1r'}$,
$D^x_{RR'} \equiv \sigma^z_{rr'} \tau^z_{2r} \tau^z_{2r'}$, and
$V^x_R \equiv \sigma^z_{12} \sigma^z_{23} \sigma^z_{34} \sigma^z_{41}$.
Compare this with the dual formulation Eq.~(\ref{dualS2ch}) of the 
two-chargon theory.
The ferromagnetic phase in the dual language corresponds to confinement
in both dual gauge fields $C$ and $D$.
In the P$^*$ phase, both these fields become deconfined while the
dual matter field $V$ where the two connect remains gapped.


\end{document}